\documentclass[useAMS,usenatbib]{mn2e}

\usepackage{epsfig,lscape} 
\usepackage{hyperref} 
\usepackage{natbib}
\usepackage[caption=false]{subfig}
\usepackage{lscape}
\usepackage{textcomp}
\usepackage{gensymb}
\usepackage{xcolor}
\usepackage{amsmath}
\usepackage{comment}
\usepackage{arydshln}
\usepackage{graphics}
\usepackage{booktabs}
\usepackage[caption=false]{subfig}

\usepackage{subfiles}

\newcommand{\mnras}{MNRAS}
\newcommand{\apj}{ApJ}
\newcommand{\araa}{ARA\&A}
\newcommand{\aap}{A\&A}
\newcommand{\aj}{AJ}
\newcommand{\apjl}{ApJL}
\newcommand{\apjs}{ApJS}
\newcommand{\pasp}{PASP}
\newcommand{\ao}{ApOpt}

\title[MW radial migration strength]{Quantifying radial migration in the Milky Way: Inefficient over short timescales but essential to the very outer disc beyond $\sim 15$~kpc}

\author[J. Lian et al.]{{Jianhui~Lian}$^{1,2}$\thanks{lian@mpia.de, gail.zasowski@gmail.com}, {Gail Zasowski}$^{1}$, {Sten Hasselquist}$^{1,3}$, {Jon Holtzman}$^{4}$, 
\newauthor Nicholas Boardman$^{1}$, Katia Cunha$^{5,6}$, Jos\'e G. Fern\'andez-Trincado$^{7,8}$, 
\newauthor Peter M. Frinchaboy$^{9}$, {D. A. Garcia-Hernandez}$^{10,11}$, 
Christian Nitschelm$^{12}$,
\newauthor Richard R. Lane$^{13}$, Daniel Thomas$^{14,15}$, Kai Zhang$^{16}$\\
\small $^{1}${Department of Physics \& Astronomy, University of Utah, Salt Lake City, UT 84112, USA}\\
\small $^{2}${Max Planck Institute for Astronomy, 69117, Heidelberg, Germany} \\
\small $^{3}${NSF Astronomy and Astrophysics Postdoctoral Fellow}\\
\small $^{4}${Department of Astronomy, New Mexico State University, Las Cruces, NM 88003, USA}\\
\small $^{5}$ {Steward Observatory, The University of Arizona, Tucson, AZ, 85719, USA} \\
\small $^{6}$ {Observat\'orio Nacional, 20921-400 So Crist\'ovao, Rio de Janeiro, RJ, Brazil}\\
\small $^{7}$ {Instituto de Astronom\'ia, Universidad Cat\'olica
del Norte, Av. Angamos 0610, Antofagasta, Chile}\\
\small $^{8}${Universidad de Atacama, Copayapu 485, Chile}\\
\small $^{9}$ {Department of Physics \& Astronomy, Texas Christian University, Fort Worth, TX 76129, USA}\\
{\small $^{10}${Instituto de Astrof\'isica de Canarias (IAC), E-38205 La Laguna, Tenerife, Spain}}\\
{\small $^{11}${Universidad de La Laguna (ULL), Departamento de Astrof\'sica, E-38206 La Laguna, Tenerife, Spain}}\\
\small $^{12}${Centro de Astronom\'ia (CITEVA), Universidad de Antofagasta, Avenida Angamos 601, Antofagasta 1270300, Chile}\\
\small $^{13}${Instituto de Astrofısica, Pontificia Universidad Cat\'olica de Chile, Av. Vicuna Mackenna 4860, 782-0436 Macul, Santiago, Chile}\\
\small $^{14}${Institute of Cosmology and Gravitation, University of Portsmouth, Burnaby Road, Portsmouth, UK, PO1 3FX}\\
\small $^{15}${School of Mathematics and Physics, University of Portsmouth, Lion Terrace, Portsmouth, UK, PO1 3HF}\\
\small $^{16}${Lawrence Berkeley National Laboratory, 1 Cyclotron Road, Berkeley, CA 94720, USA}
}

\begin{document}

\maketitle

\begin{abstract}
    Stellar radial migration plays an important role in reshaping a galaxy's structure and the radial distribution of stellar population properties. 
    In this work, we revisit reported observational evidence for radial migration and quantify its strength using the age--[Fe/H] distribution of stars across the Milky Way with APOGEE data. 
    {We find a broken age-[Fe/H] relation in the Galactic disc at $r>6$~kpc,} with a more pronounced break at larger radii.
    To quantify the strength of radial migration, we assume stars born at each radius have a unique age and metallicity, and then decompose the metallicity distribution function (MDF) of mono-age young populations into different Gaussian components that originated from various birth radii at $r_{\rm birth}<13$~kpc. 
    We find that, at ages of 2 and 3~Gyr, roughly half the stars were formed within 1~kpc of their present radius, 
    {and very} few stars ($<5$\%) were formed more than 4~kpc away from their present radius. These results suggest limited short distance radial migration and inefficient long distance migration in the Milky Way during the last {3} Gyr. 
    In the {very} outer disc beyond 15~kpc, the observed age--[Fe/H] distribution is consistent with the prediction of pure radial migration from smaller radii, suggesting a migration origin of the very outer disc. 
    We also estimate intrinsic metallicity gradients at ages of 2 and 3~Gyr of $-0.061$~dex~kpc$^{-1}$ and $-0.063$~dex~kpc$^{-1}$, respectively. 
\end{abstract}

\section{Introduction}
The radius of a disc star's orbit can vary substantially over time when subjected to perturbations, a general process usually referred to as radial migration \citep[e.g.,][]{lynden-bell1972,sellwood2002}. There are two dominant modes of this orbital variation: 
change of a star's guiding radius (i.e., change of orbit angular momentum) and change of oscillation amplitude around the orbital guiding radius, which are dubbed churning or cold torquing, and blurring or kinematic heating, respectively \citep{sellwood2002,daniel2019}. 
Theoretical and numerical studies suggest that a star might migrate away from its birth guiding radius for a variety of reasons, including 
non-resonance interactions with an infalling satellite galaxy \citep{quillen2009} or
giant molecular clouds in the disc \citep{schonrich2009}, or resonance interactions with non-axisymmetric patterns like the bar (e.g., \citealt{brunetti2011,kubryk2013,dimatteo2013,halle2015,khoperskov2020}), spiral arms (e.g., \citealt{sellwood2002,roskar2008a,loebman2016}) or overlap between the two \citep{minchev2010,daniel2019}. 

While radial mixing is commonly seen in numerical simulations (e.g., \citealt{roskar2008a,dimatteo2013,grand2015,elbadry2016,buck2020,vincenzo2020}), unambiguous observational evidence for radial migration is limited. One of the major effects of radial migration is mixing stars born at different radii with potentially different chemical abundances, which results in a highly complex age--metallicity distribution at a single present-day radius. Many reported lines of observational evidence for radial migration are therefore found in the age--chemistry plane. 
For example, \citet{haywood2008} found high metallicity dispersion in the age--metallicity distribution at the solar neighborhood and that the dispersion increases with age, which can both be explained by radial migration \citep{roskar2008b}. Such high dispersion in age--metallicity distribution is also seen in other stellar spectroscopic surveys \citep[e.g.,][]{bergemann2014,anders2017,xiang2017,lin2018}. 
A broader radial profile of older populations found in \citet{mackereth2017} is also a possible signal of radial migration. 
Another recently discovered feature in the age--metallicity distribution {that is possibly caused by radial migration} is the younger age of solar metallicity stars compared to stars with super-solar metallicity \citep{anders2017,feuillet2018,silva2018,hasselquist2019,lian2020a}, {which suggests} an interrupted age--metallicity relation.  
This feature is further confirmed with different observations from LAMOST survey \citep{xiang2017,wu2018}. 
{A positive age gradient is observed in the outskirts of local galaxies, as opposed to a more common negative age gradient in the inner regions \citep{roskar2008a,bakos2008,yoachim2012,ruiz-lara2017}, which is also belived to be a possible observational signature of radial migration.} Whether such a positive age gradient is present in the outer disc of the Milky Way is still unclear and will be explored in this paper.  

In addition to the features in the age-metallicity distribution and a positive age gradient, additional possible observational evidence for radial migration comes in the form of the radially variant metallicity distribution function (MDF), thanks to the advent of numerous massive stellar spectroscopic surveys that map a large portion of the Galaxy (e.g., LAMOST, APOGEE, GALAH). \citet{hayden2015} studied the MDF at different radial and vertical positions in the disc and found that the MDF of the low-$\alpha$ population in the disc plane shows clear radial variation, with negative skewness in the inner disc and positive skewness in the outer disc beyond the solar radius. They further showed that this change of skewness can be explained by radial migration alone. 

{Attempts have been made to build connections between some of the observational features mentioned above and radial migration in theoretical models. When combined with a radially variant {monotonic} chemical enrichment history across the disc, radial migration is believed to be one of the possible explanations for {the observed complex age-distribution.} }
In this picture, the relatively old, metal-rich stars at the solar neighbourhood originated from the inner Galaxy where the early enrichment process was more efficient, while the younger, lower-metallicity stars either formed locally or migrated from the outer disc (e.g., \citealt{minchev2013}). 
However, another scenario that invokes a later, metal-poor gas accretion event has also been shown to be able to explain the complex age-metallicity distribution with an interrupted age-metallicity relation \citep[e.g.,][]{spitoni2019,spitoni2020,lian2020a,lian2020b,renaud2021}. 
Such a delayed gas accretion has long been considered in the literature to explain the stellar evolution track in [$\alpha$/Fe]--[Fe/H] in the solar neighborhood \citep[e.g.,][]{chiappini1997,calura2009,haywood2019,spitoni2019}. 
Radial migration is also shown to be responsible for the radially variant MDF, as illustrated
in \citet{loebman2016} using galaxy dynamical simulations, \citet{frankel2018} and \citet{sharma2020} with analytical models, and \citet{johnson2021} which used a hybrid model that combines the two. \citet{frankel2018} further derived the radial migration strength by fitting their empirical model to the observed age--[Fe/H] distribution of red clump stars across the disc. One caveat in these empirical models is that {a relatively simple star formation history is generally assumed} while the MDFs under consideration comprise stars formed over a long period of time, such that the shape of the MDF is very sensitive to the local star formation and chemical enrichment history as well \citep{johnson2021}. A possibility that a scenario with a radially variant {complex multi-phase SFH without radial migration} can explain the radial change of the MDF shape cannot be excluded (Lian et al. in prep).  

In this paper we revisit the reported observational signatures of radial migration in age--chemistry space and explore more direct and stringent observational constraints using the latest observations from the APOGEE survey \citep{majewski2017}. Among all stellar chemistry surveys to date in the Milky Way, APOGEE provides the most comprehensive coverage in radius, particularly at low vertical height. This enables us for the first time to inspect the radial trends of stellar age and chemical abundance, with a homogeneous dataset, continuously from the bulge region to the outer disc (as far as $\sim$20~kpc from the Galactic center). In addition to the wide radial coverage, another improvement in this work is that we restrict the MDF analysis to mono-age populations that exhibit relatively steep radial metallicity gradient at formation, which minimizes the effect of star formation history. 

This paper is organised as following: we briefly introduce the data and sample selection in \textsection2, and in \textsection3 we present the distribution of our sample in age--[Fe/H] space and discuss its behavior with radius. 
In \textsection4 we describe the quantitative constraints on the radial migration strength obtained by performing a detailed analysis of mono-age MDFs across the Galaxy. We then compare our results with previous works and discuss potential biases in our analysis in \textsection5. Finally, a brief summary is included in \textsection6.     

\section{Sample Selection}
The stellar sample in this work is selected from APOGEE observations contained in SDSS-IV Data Release 16 \citep[DR16;][]{ahumada2020,jonsson2020}, plus a post-DR16 APOGEE internal data release that includes data from observations through March 2020, reduced with a very slightly updated version of the DR16 pipeline (r13)\footnote{The allStar catalog used is allStar-r13-l33-58932beta.fits.}. APOGEE is a near-infrared, high-resolution spectroscopic survey \citep{blanton2017,majewski2017} that primarily targets evolved giant stars in the Milky Way and Local Group satellites \citep[][Santana et al. in prep]{zasowski2013,zasowski2017,beaton2021}. This survey is performed using custom spectrographs \citep{wilson2019} with  the  2.5~m  Sloan  Telescope  and  the  NMSU 1~m Telescope at the Apache Point Observatory \citep{gunn2006,holtzman2010}, and with the 2.5~m Ir\'en\'ee du~Pont telescope at Las Campanas Observatory \citep{bowen1973}. 

We use chemical abundances ([Fe/H], [Mg/Fe]) and stellar parameters (i.e., log$(g)$ and $T_{\rm eff}$) derived by custom pipelines described in \citet{nidever2015} and \citet{garcia2016} and line list in \citet{smith2021}, and spectro-photometric distances based on the procedure described in \citet{rojas2017}. We note that the usage of [Fe/H] instead of [M/H] implicitly excludes some cool ($T_{\rm eff}<$4000~K) metal-rich ([Fe/H]$>$0.1) stars which have [Fe/H] and [M/H] measurements differ greater than 0.1 and therefore do not have [Fe/H] populated in the released catalog. Since these stars mostly have intermediate ages (ages$>$5~Gyr) which is not the main sample of the analysis here, the results of this paper would not be affected significantly by this selection effect.  
As we focus on the radial migration caused by churning (i.e., a change in guiding radius), we use guiding radius ($r_{\rm guide}$, the radius of a circular orbit with equivalent angular momentum) instead of the Galactocentric radius ($r_{\rm GC}$) to represent the position of a star in the Galaxy. The guiding radius is obtained by integrating orbits using the {\sl galpy} Python package with the MWPotential2014 model for Milky Way gravitational potential \citep{bovy2015,rojas2020}, adopting our distances, APOGEE radial velocities, and {\it Gaia} DR2 proper motions.   

We use the recommended stellar ages in the DR16 astroNN Value Added Catalog \citep{mackereth2019}\footnote{https://data.sdss.org/sas/dr16/apogee/vac/apogee-astronn}, which are derived by training a Bayesian neural network model \citep{leung2019}\footnote{https://github.com/henrysky/astroNN} with asteroseismic ages derived from {\sl Kepler} and APOGEE combined observations \citep{pinsonneault2018}. 
We note that astroNN spectroscopic ages rely on the correlation between [C/N] abundance ratio and stellar age for red giant stars. However, for low-gravity, low-metallicity (log($g)<2$, [Fe/H]$<-$0.4) giants, the surface [C/N] is further affected by extra mixing after the first dredge-up, with a stronger effect at lower gravity and lower metallicity \citep{shetrone2019}. For this reason, the spectroscopic ages of low-gravity, low-metallicity stars based on [C/N], without explicitly considering this extra mixing, should be used with caution. 

According to \citet{shetrone2019}, for $\alpha$-rich stars, the effects of extra mixing start to appear at $\rm [Fe/H] \sim -0.4$. As we will show in \textsection3, the age--[Fe/H] relation of our high-$\alpha$ stars shows a break at $\rm [Fe/H] \sim -0.5$, with younger ages for stars with $\rm [Fe/H] < -0.5$. This unphysical trend is likely due to the lack of considering extra mixing effects in the age determination of metal-poor stars. Unlike the high-$\alpha$ stars, however, the age--[Fe/H] relation of low-$\alpha$ stars extends consistently, without any break, to $\rm [Fe/H] \sim -0.7$. 
We also confirm that the high- and low-gravity (log($g$)$>$2.7 and log($g$)$<$1) low-$\alpha$ stars follow the same age--[Fe/H] relation at a given radius, suggesting no dependence of age on surface gravity. These findings indicate that the extra-mixing effect may occur at lower [Fe/H] for low-$\alpha$ stars. 
{We note that the training set of the astroNN neural network does not include low-$\alpha$ stars at $\rm [Fe/H]<-0.5$. The astroNN ages of these stars are obtained by extrapolation from more metal-rich stars and could potentially be subject to unaccounted-for systematic errors. In this paper, we use the astroNN-derived ages for the whole of our low-$\alpha$ sample, which extends down to $\rm [Fe/H] \sim -0.7$. 
We recommend that readers treat the results of the outer disc (\textsection4.2) as preliminary, pending future confirmation with updated ages that are calibrated to asteroseismic observations of stars with $\rm [Fe/H]<-0.5$.} 
    

To select a sample of giant stars with reliable measurements, following our previous works (e.g., \citealt{lian2021}), we apply the following selection criteria to the parent APOGEE catalog: 
\begin{itemize}
  \item Signal-to-Noise ratio (SNR) $>$ 70,
  \item $T_{\rm eff} > 3500$~K,
  \item $\log{(g)} < 3.3$,
  \item Vertical height $|z| < 2$~kpc,
  \item Orbital eccentricity less than 0.5,
  \item $\rm EXTRATARG==0$, 
  \item APOGEE\_TARGET1 and APOGEE2\_TARGET1 bit 9 $== 0$,
  \item STARFLAG bits 4, 9, 16, and $17 == 0$, and
  \item ASPCAPFLAG bits 19 and $23 == 0$, no flag for BAD metals and BAD overall for star. 
\end{itemize}
The EXTRATARG bitmask indicates a number of targeting considerations; EXTRATARG==0 identifies main survey stars and removes duplicated observations. APOGEE\_TARGET1 or APOGEE2\_TARGET1 bit 9 are set for targets that are possible star cluster members. The STARFLAG bitmask describes things worth noting in the spectrum, and the ASPCAPFLAG includes all kinds of warnings in the determination of stellar parameters.   
For more detailed descriptions of the APOGEE bitmasks, we refer the reader to \url{https://www.sdss.org/dr16/algorithms/bitmasks/}.
Our final sample contains 232,166 stars. We discuss potential selection biases of the stellar parameter criteria in \textsection4.1. 

\section{Observed age--[Fe/H] distribution across the Galaxy}

{In this section, we present a short empirical overview of the data, to highlight the qualitative patterns and trends that guided our assumptions and methods used in the later quantitative analyses.}

\begin{figure*}
	\centering
	\includegraphics[width=18cm,viewport=140 70 1400 980,clip]{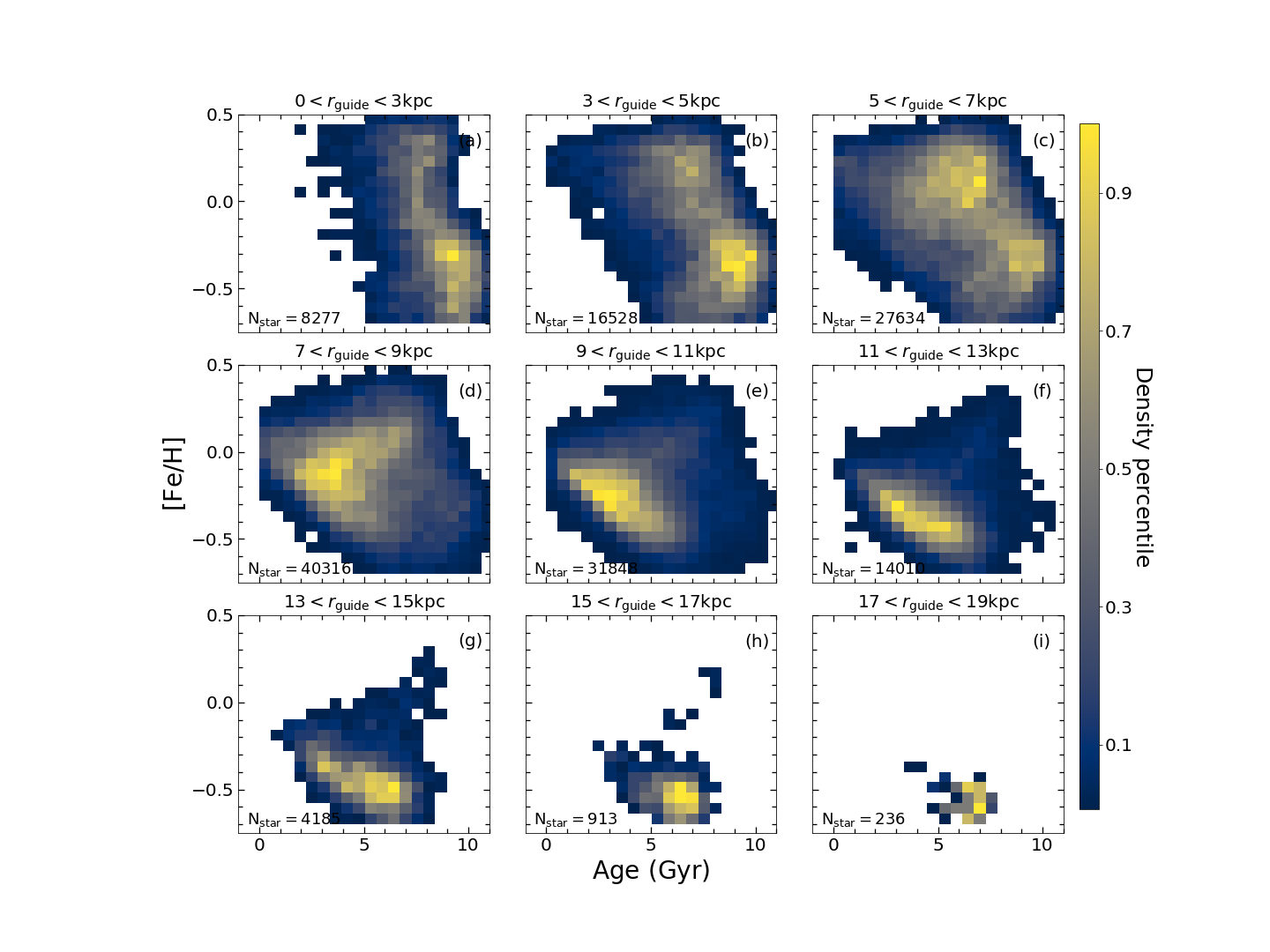}
	\caption{Age--[Fe/H] distribution as a function of guiding radius, from top left (panel a) for the bulge region to the bottom right (panel i) for the {very} outer disc. Note the presence of two dominant age--[Fe/H] sequences, with complementary age coverage at intermediate radii (e.g., panel d). 
	} 
	\label{age-feh-radius}
\end{figure*} 

\begin{figure*}
	\centering
	\includegraphics[width=16cm,viewport=10 10 1100 500,clip]{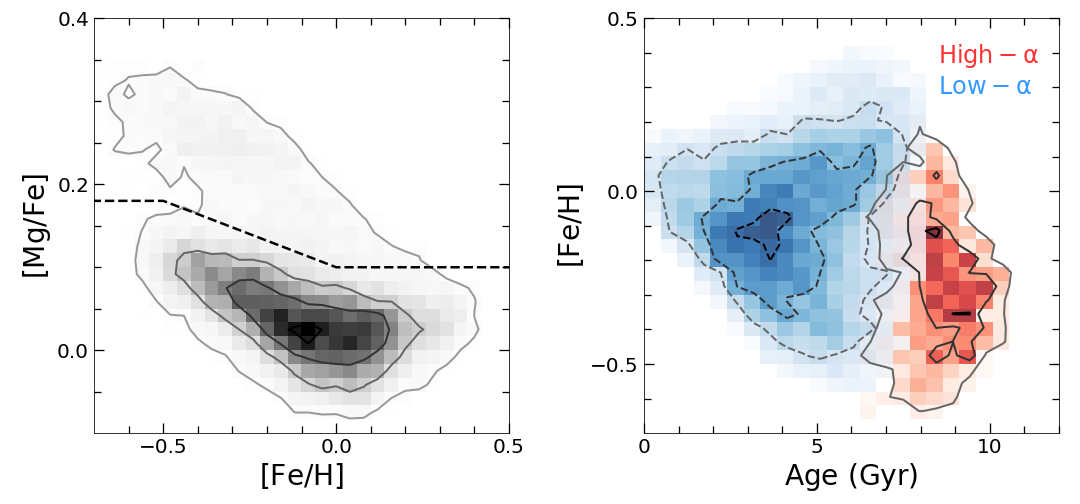}
	\caption{{\sl Left}: [Mg/Fe]--[Fe/H] distribution at the solar radius ($7<r_{\rm guide}<9$~kpc). The gray shading indicates the stellar density, with gray contours at 5\%, 30\%, 60\%, and 90\% of the peak density. The dashed line denotes the demarcation line between the high- and low-$\alpha$ branches \citep[adapted from][]{adibekyan2011}. {\sl Right}: Age--[Fe/H] distribution of the stars shown in the left panel, separated into the high- and low-$\alpha$ subsamples (colored in red and blue, respectively). Each set of density contours indicates 30\%, 60\%, 90\% of the respective peak density. 
	} 
	\label{age-feh-mgfe}
\end{figure*} 


Figure~\ref{age-feh-radius} shows the stellar distributions in the age--[Fe/H] plane at different guiding radii ($\Delta r_{\rm guide} = 2$~kpc, except for the innermost bin with $\Delta r_{\rm guide} = 3$~kpc). 
The age--[Fe/H] distribution varies dramatically with radius.  
In the inner Galaxy (top row: panels a, b, and c), most stars are older than 5~Gyr and fall along a clear age--[Fe/H] relation that occupies the relatively old, metal-rich quadrant in the age--[Fe/H] plane. {The nearly vertical distribution in the first panel of Fig.~\ref{age-feh-radius} reflects rapid chemical enrichment in the bulge.} Stars in the outer disc, beyond the solar radius (panels e through i), follow an age--[Fe/H] relation distinct from the inner Galaxy, with systematically younger age and lower [Fe/H]. 
Near the solar radius (panel d), both of these age--[Fe/H] sequences contribute to a complex age--[Fe/H] distribution. This complex pattern in the solar neighborhood has been seen in many recent works, based on different datasets and different age determination methodologies {(e.g., \citet{nissen2020}; \citet{jofre2021}; Fig.~3 in \citealt{feuillet2019}; Fig.~20 in \citealt{wu2019}; and Fig.~6 in \citealt{sahlholdt2021}). 

These two age--[Fe/H] sequences are related, but not identical, to the two branches in the [$\alpha$/Fe]--[Fe/H] plane. 
Figure~\ref{age-feh-mgfe} shows the distribution of solar neighborhood stars ($7<r_{\rm guide}<9$~kpc) in [Mg/Fe]--[Fe/H] (left-hand panel) and in age--[Fe/H] (right-hand panel, c.p.\ to Figure~\ref{age-feh-radius}d). We adopt the demarcation between the high- and low-$\alpha$ branches from \citet{lian2020b}, which was adapted to APOGEE observations from the separation curve in \citet{adibekyan2011}. It can be seen in the right panel of Figure~\ref{age-feh-mgfe} that the high-$\alpha$ population occupies the lower right portion of the {older} age--[Fe/H] sequence, while the low-$\alpha$ stars span both the entire {younger} age--[Fe/H] sequence {\it and} the most metal-rich portion of the {older} sequence. 
In comparison, the bulge's stars, though lying along a single single age--[Fe/H] sequence (Figure~\ref{age-feh-radius}a), show similar gradients in [$\alpha$/Fe]; gradients are also seen along this sequence in kinematics, the old metal-poor stars having higher velocity dispersion and the younger metal-rich stars having colder, more bar-like kinematics \citep[e.g.,][]{babusiaux2010,hill2011,schultheis2017,rojas2019,queiroz2020}.

\begin{figure}
	\centering
	\includegraphics[width=16.5cm,viewport=10 10 1100 500,clip]{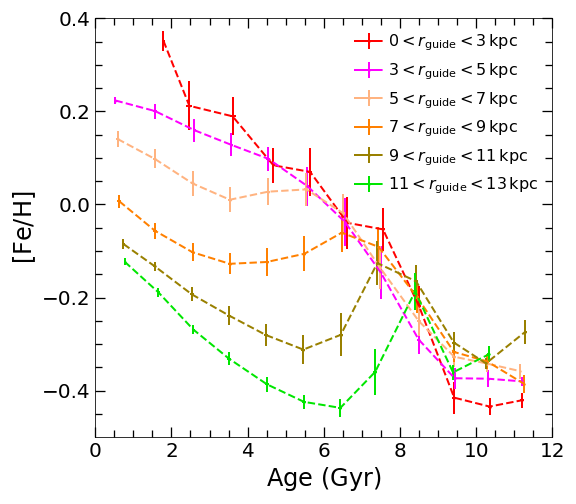}
	\caption{{Median [Fe/H] as a function age at different guiding radii.} Error bars indicate the uncertainty of the median [Fe/H] at a given age, estimated by bootstrapping. 
	} 
	\label{age-feh-median}
\end{figure}

The radial variations in both age--[Fe/H] and [$\alpha$/Fe]--[Fe/H] are signatures of radial variations in the Milky Way disc's star formation history and of subsequent rearrangement of the stars.
{Prior to the more extensive modeling in later sections, we show here a simple experiment of
disentangling the ex-situ and in-situ scenarios by comparing the position of stars in the age--[Fe/H] plane located at different Galactic radii.} {\it The basic idea is that stars born at the same radius but currently located at different radii should still follow the same age--[Fe/H] relation.}

Figure~\ref{age-feh-median} shows the median stellar metallicity as a function of age and guiding radius. 
The error bars indicate the uncertainty in the median [Fe/H] at a given age bin, estimated via bootstrapping. The median age--[Fe/H] relations at $r_{\rm guide}=6$~kpc and beyond deviate from a single, monotonic age-[Fe/H] relation and have broken profiles, possibly as a result of dilution induced by a late metal-poor gas accretion event \citep{haywood2019,spitoni2019,lian2020a,lian2020b}. At ages $<6$~Gyr, stars at smaller radii have systematically higher [Fe/H], indicating a negative metallicity gradient {not (yet) erased by significant radial mixing in these young stars. This is one basis for the assumption that the majority of young stars observed at each present radius formed locally and reflect the recent chemical gradient of the ISM \citep[e.g.,][]{minchev2018}.} At age $>6$~Gyr, however, the radial variation of [Fe/H] becomes negligible at most radii, and we cannot exclude a non-local (i.e., migration) origin for these stars based on chemistry alone\footnote{It is interesting to note that, given an age of 4.6~Gyr \citep{bonanno2002} and [Fe/H] = 0, the position of the Sun in age-[Fe/H] plane is best consistent with the disc at $5<r_{\rm guide}<7$~kpc, suggesting that the Sun was likely born at smaller radii and migrated outward to the present location ($r_{\rm \odot}=8.2$~kpc, \citealt{bland2016}) which is qualitatively consistent with previous result in \citet{minchev2018}.}.



Unlike the majority of the disc, the outermost radial bins (from 13 to 19~kpc; panels g, h, and i in Fig.\ref{age-feh-radius}) show a pattern in which the age--[Fe/H] distribution is more concentrated towards lower [Fe/H] and older age as radius increases. This trend is unlikely to be caused by a lack of disc plane coverage at larger radii or a selection bias with age. The APOGEE main survey targets stars with Galactic coordinates on a semi-regular grid with preference on the disc plane \citep{zasowski2013,zasowski2017}. {We also inspect the age-[Fe/H] distribution at different radii after resampling the $\log(g)-T_{\rm eff}$ distribution to that of the outermost radial bin at $17<r_{\rm guide}<19$~kpc and confirm the existence of this trend.} 
The interesting behavior of age-[Fe/H] distribution} suggests that the Milky Way also likely exhibits a positive gradient of mean age in the {very} outer {disc}, similar to the pattern found in other galaxies that has been argued to be a signature of radial migration \citep{ruiz-lara2017}. 

{Based on the patterns described here, in the following sections} we will analyze the age--[Fe/H] distribution in the regions with $r_{\rm guide}<$13~kpc (\textsection{4.1}) and with $r_{\rm guide}>$13~kpc (\textsection{4.2}) separately.  

\section{Results}
\subsection{Quantifying radial migration within $r<13$~kpc}
In this section we quantitatively measure the strength of radial migration in the Milky Way by analyzing the MDF of {\sl mono-age} populations. 
The approach is based on the findings that the young stars in the Galaxy formed at different radii have significantly different [Fe/H] (\textsection{3.2}), 
which allows us to use the [Fe/H] of a young star to trace its birth radius. A similar approach was used by \citet{minchev2018} to recover the birth radius distribution of solar neighborhood stars and the evolution of metallicity gradient.  
Once the birth radius distribution of stars at a given present position is known, we can quantify the relative contribution of {local} formation and radial migration at that position in the Milky Way's disc. Since the stars in the first age-[Fe/H] sequence {have indistinguishable [Fe/H] at different radii for a given age as discussed in \textsection 3.2}, this approach is not applicable to infer the radial migration effect at the early evolution stage of the Milky Way.

\subsubsection{Mono-age MDFs}

\begin{figure*}
	\centering
	\includegraphics[width=18cm,viewport=10 10 1400 420,clip]{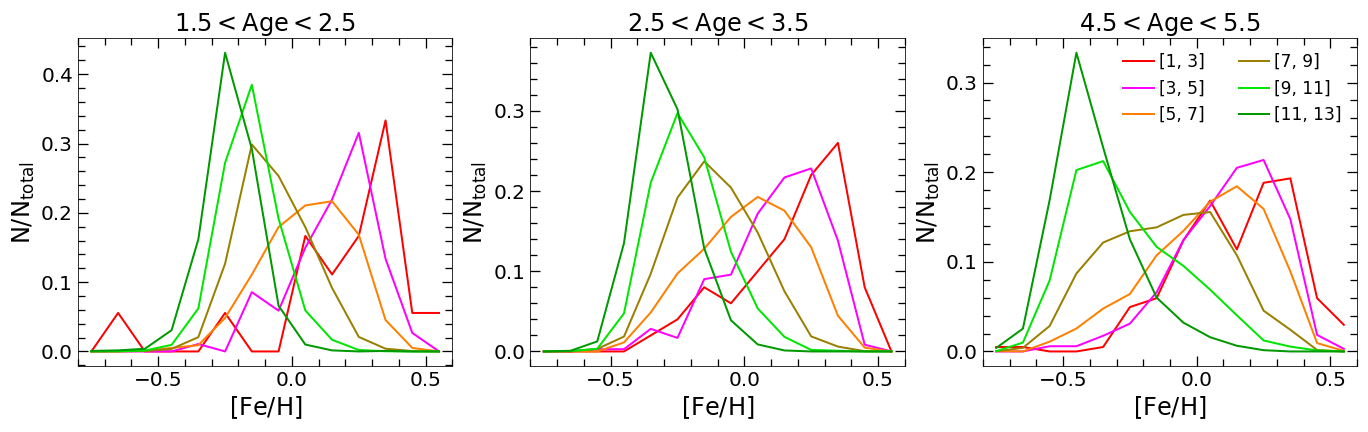}
	\caption{Normalized metallicity distribution functions of stars in age bins of $\Delta\tau=1$~Gyr, at ages of 2~Gyr (left), 3~Gyr (middle), and 5~Gyr (right). Each panel contains one mono-age group of stars sorted by guiding radius, denoted by colors as in the legend in the right panel. {The clearly skewed and broadened mono-age MDFs within solar radius are strong evidence for radial migration, while the narrow symmetric MDF at $11<r_{\rm guide}<13$~kpc indicates least contamination by radial migration at this radius.}
	} 
	\label{mdf-ages}
\end{figure*} 

Figure~\ref{mdf-ages} shows the MDF of stars with ages around 2~Gyr (left), 3~Gyr (middle), and 5~Gyr (right), with a width of 1 Gyr for each age bin. The colored lines in each panel correspond to stars of that age in different bins of guiding radius, as indicated by the legend in the right-hand panel. The stars at 5~Gyr are used to demonstrate the upper age limit beyond which the mono-age MDF analysis conducted in this work is not applicable. 

The mono-age MDFs vary dramatically with guiding radius, in both peak metallicity and MDF shape. The peak [Fe/H] of the mono-age MDFs shifts systematically towards lower values at larger radii. 
In the outer disc, the MDFs of the mono-age stars are single-peaked Gaussian profiles, with relatively narrow distributions. At intermediate radial bins, such as the solar radius, the MDFs are much broader, indicative of some level of mixing due to radial migration. 

In the inner Galaxy, the mono-age MDFs exhibit clear negative skewness, with extended tails towards low [Fe/H]. 
These broad, skewed mono-age MDFs cannot be easily explained by chemical evolution models without invoking, for example, tuned and inhomogeneous star formation. 
While fine-tuning the ISM metallicity gradient or the strength of inhomogeneous (stochastic and/or azimuthally-varying) star formation as a function of radius might be able to explain these observations, a more natural explanation is radial migration combined with radially-varying SFHs that mixes stars born at the same time but at different radii with different chemical abundances. 
The strength of local stochastic star formation is still unclear. For the variation in azimuthal direction, a mixture of results have been reported in the literature. 
Some works have found no significant large-scale azimuthal variation in the metallicity of gas and young stars in the Milky Way (e.g., \citealt{luck2011,bovy2014}) and other disc galaxies (e.g., \citealt{li2013,kreckel2020}), while many other studies reported differences in the ISM and stellar metallicity gradient along different azimuth angles (Milky Way: \citealt{davies2009,balser2015,wenger2019}; other galaxies: \citealt{sanchez2015,ho2017}). 

\subsubsection{Mono-age MDF decomposition}
\begin{figure*}
	\centering
	\includegraphics[width=18cm,viewport=10 10 1400 860,clip]{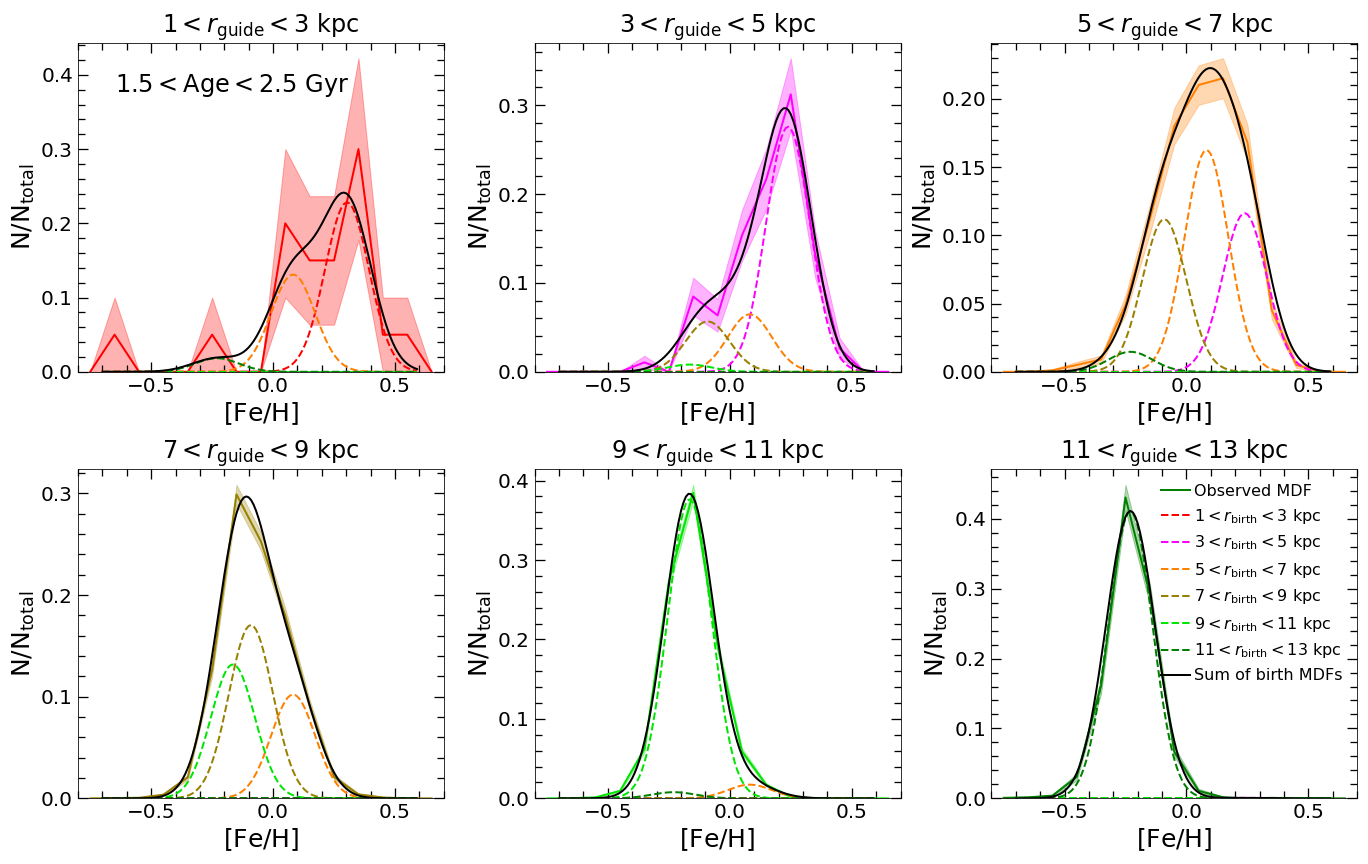}
	\caption{Decomposition of mono-age MDFs (age of 2~Gyr) at different guiding radii. Each panel indicates one radial bin, with that bin's observed MDF denoted by the solid coloured line. {Shaded region indicates uncertainty of the distribution, assuming Poisson noise.} The dashed colourful lines are the best-fitted {birth} MDFs from different birth radii {as illustrated in the bottom right legend} and the black line is the sum of these {birth} MDFs (see \textsection4.1.2). The colour scheme is the same as in Fig.~\ref{mdf-ages}. 
	} 
	\label{mdf-decomp-age2}
\end{figure*} 

\begin{figure*}
	\centering
	\includegraphics[width=18cm,viewport=10 10 1400 860,clip]{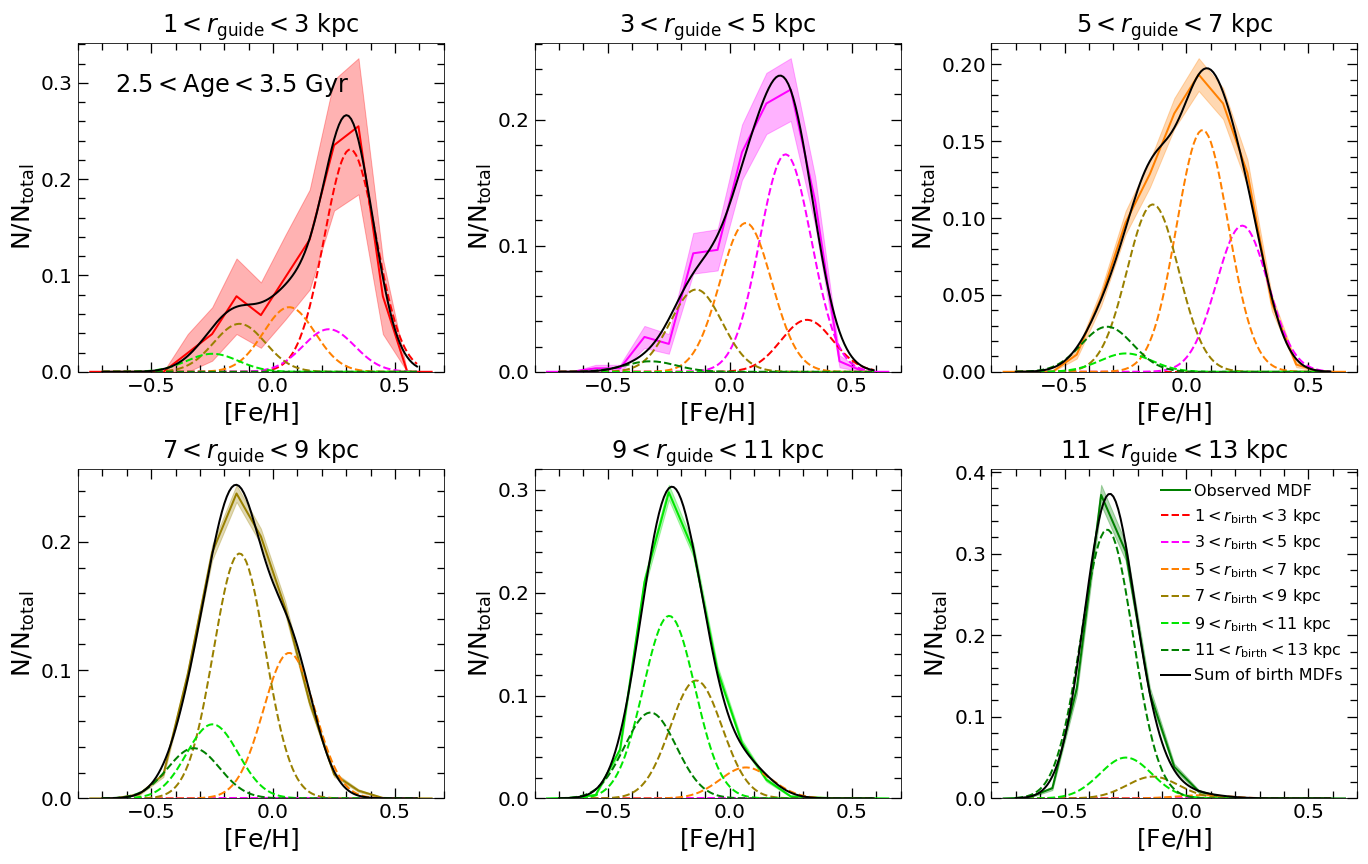}
	\caption{Similar to Fig.~\ref{mdf-decomp-age2} but for mono-age MDFs at age of 3~Gyr. 
	} 
	\label{mdf-decomp-age3}
\end{figure*} 

\begin{figure*}
	\centering
	\includegraphics[width=18cm,viewport=10 10 1400 860,clip]{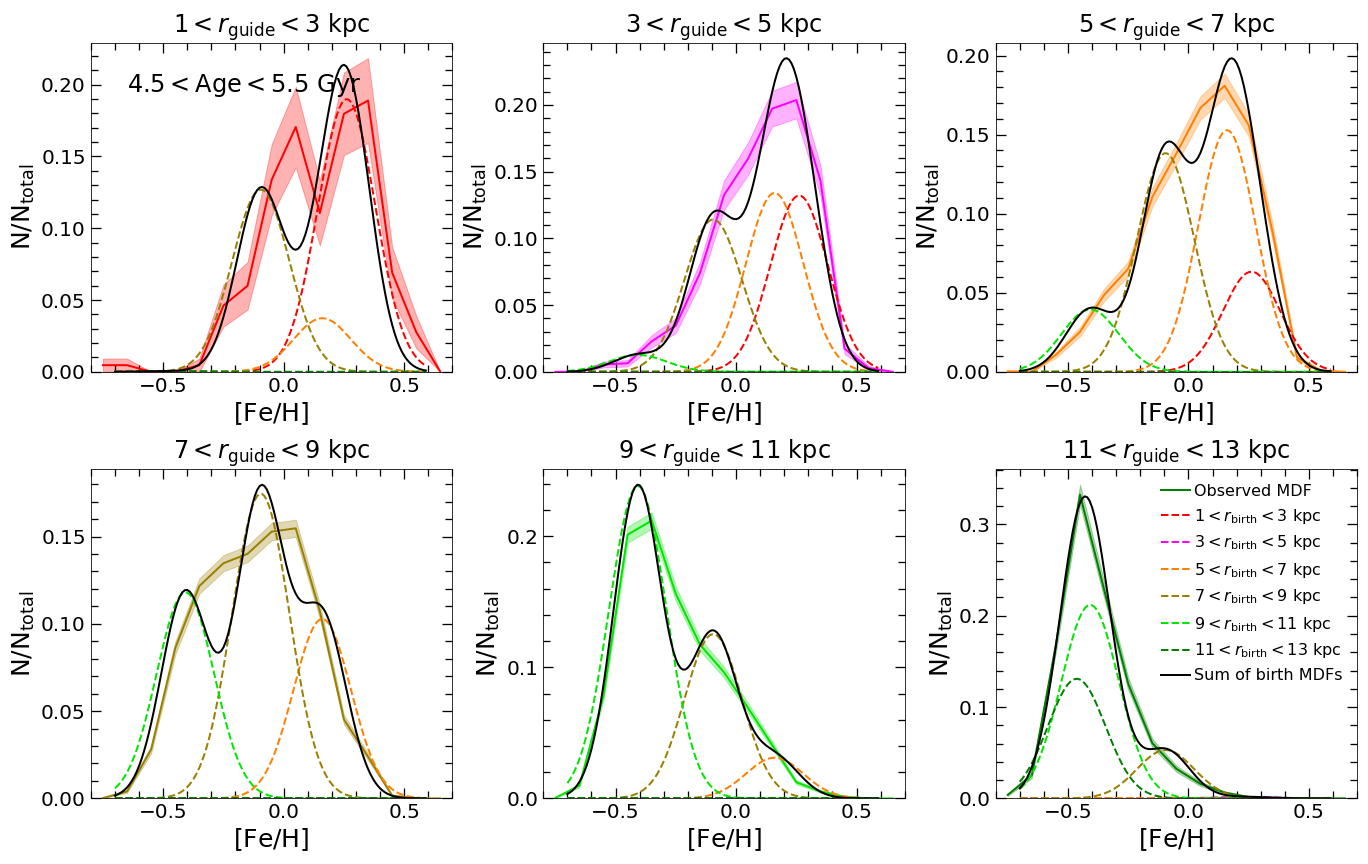}
	\caption{Similar to Fig.~\ref{mdf-decomp-age2} but for mono-age MDFs at age of 5~Gyr. 
	} 
	\label{mdf-decomp-age5}
\end{figure*} 

To identify migrated stars and their contribution to the observed MDF at their present radius, we need to understand the {birth} MDF of locally formed stars in each age and radial bin. Here we make a small number of important assumptions about the {birth} MDFs of mono-age populations, {including: 
\begin{itemize}
 \item a Gaussian profile for the birth MDFs,
 \item same width of the birth MDFs at various radii,
 \item the peak of the observed MDF the same as the that of the birth MDF.
\end{itemize}
}

The first assumption {we make} is that the mono-age MDF of {locally formed} stars has a Gaussian profile. 
In principle, at a given spatial position and moment in time, new stars should share the same [Fe/H] and elemental abundance pattern. However, in practise, the observational uncertainties in both Galactic radius and stellar age would broaden the observed MDF of even mono-age {populations}, with the width of broadening depending on the magnitude of observational uncertainties, the metallicity gradient, inhomogeneity in the local star formation, and the size of the age/radial bins chosen. 
We test this assumption with a mock stellar sample from a simple chemical evolution model \citep{lian2020c}.
The MDF of the mock stars in a narrow age bin indeed presents a Gaussian-like profile. {As no evidence for strong dependence of observational errors on radius has been reported for APOGEE} and the metallicity gradient is generally linear (e.g., \citealt{anders2017}), given the same size of radial bins used, we assume the width of the Gaussian MDF does not change with radius. Note that this assumption also relies on a precondition that azimuthal variations of chemical enrichment do not change with radius, which has not been extensively studied and needs to be confirmed with further studies. Since the outer disc presents the narrowest mono-age MDFs, suggesting least contamination from radial migration, we fit the observed MDF at $11<r_{\rm guide}<13$ with a Gaussian profile and consider the width to represent the mono-age {birth} MDFs at other radial bins. The fitted 1$\sigma$ widths for the three age bins at 2, 3, and 5~Gyr are 0.088, 0.104, 0.119~dex, respectively. The wider {birth} MDF at older ages is likely due to a steeper intrinsic metallicity gradient of older stellar populations (see more discussion in \textsection{5.3}). We also assume azimuthal symmetry in the SFH, such that birth spatial location can be collapsed to a 1-D radial coordinate. This assumption is supported by some recent observational results that suggest insignificant azimuthal variations in chemical abundances in the Milky Way (\citealt{luck2011}; \citealt{bovy2014}, {but see a different result in \citealt{davies2009,balser2015,wenger2019}}).  

One additional critical assumption we make is that the peak of the {birth} MDF at each radius is the same as the peak of the observed MDF of stars in that age bin. This assumption is based on the fact that our young stars that formed at different radii have disparate [Fe/H]. Therefore, at any present radius, stars migrated from other radii will have different [Fe/H] from {locally formed} stars and thus broaden and skew the full observed MDF. As long as the migrated population from a single radial bin are not a large fraction of stars in the local radial bin, the mode of the observed MDF will not be significantly affected. As discussed above, the significant radial variation of the age--[Fe/H] {relation at age$<6$~Gyr} implies the radial migration effect, if present, is not prominent. Some simulation works also suggest that stars that have migrated more than 2~kpc away from the birth radius comprise only 25\% of the whole disc stellar population \citep[e.g.,][]{roskar2008b,loebman2016}. 


Given these assumptions, {we decompose the mono-age MDFs in a sum of 6 Gaussians representing the {birth} MDFs spanning $1<r_{\rm birth}<13$.}
To estimate the peak position, we fit each observed MDF with a skewed Gaussian profile; the fitted peak [Fe/H] values are listed in Table~\ref{tab:peakz}. 
As described above, the width of each {birth} MDF is fixed to be the same as that of the outer radial bin ($11<r_{\rm guide}<13~$kpc) at the given age. For each component, then, only its amplitude is a free parameter. Therefore, to decompose each observed mono-age MDF, we have six free parameters corresponding to the amplitudes of the six radial components considered. {At each radius and age bin, the MDF decomposition can be described as}
\begin{equation*}
    f_{\rm observed}({\rm [Fe/H]}|R_{\rm guide},age) = \sum_{i=1}^{6} \alpha_{i}f_{\rm birth}({\rm [Fe/H]}|R_{{\rm birth},i},age),
\end{equation*}
{with the mode in each radial bin $i$ empirically determined in the present-day data. We fit for the $\alpha_i$ (where $\sum_{i=1}^{6}\alpha_i$=1). }

\begin{table}
    \caption{Peak [Fe/H] of the observed MDF in each age and radial bin.}
    \centering
    \begin{tabular}{l c c c}
    \hline\hline 
    Radial bins & Age$=$2~Gyr & Age$=$3~Gyr & Age$=$5~Gyr \\
    \hline
    $r_{\rm guide}$=[1 ,3] & 0.312$\pm$0.064 & 0.318$\pm$0.038 & 0.261$\pm$0.072 \\
$r_{\rm guide}$=[3 ,5] & 0.233$\pm$0.032 & 0.233$\pm$0.023 & 0.231$\pm$0.019 \\ 
$r_{\rm guide}$=[5 ,7] & 0.083$\pm$0.023 & 0.067$\pm$0.017 & 0.158$\pm$0.014 \\
$r_{\rm guide}$=[7 ,9] & -0.095$\pm$0.017 & -0.143$\pm$0.007 & -0.097$\pm$0.018 \\ 
$r_{\rm guide}$=[9 ,11] & -0.168$\pm$0.003 & -0.252$\pm$0.005 & -0.415$\pm$0.016 \\ 
$r_{\rm guide}$=[11 ,13] & -0.233$\pm$0.005 & -0.328$\pm$0.005 & -0.466$\pm$0.024 \\
    \hline
    \end{tabular}
    \label{tab:peakz}
\end{table}

Figure~\ref{mdf-decomp-age2} shows the best-fitted result for the age$=$2~Gyr MDFs in our six radial bins. The solid coloured line in each panel is the observed MDF at each present radius, along with a shaded region indicating Poisson errors. The dashed coloured lines are the best-fitted {birth} MDFs from each birth radius (with matching colours) and the black line is the sum of these {birth} MDFs. The multi-component fitted MDFs match well the observed ones at each present radius, with the largest contribution coming from the stars that formed locally. The noisy observed MDF at the innermost radial bin ($1<r_{\rm guide}<3$~kpc) is due to the small number of stars there in this age range. {The age distributions of both super- and sub-solar metallicity stars at $1<r_{\rm guide}<3$~kpc show clear extended tails at young ages, suggesting that these young stars are not likely scatter from old populations given symmetric log(age) uncertainties.} The presence of a small fraction of young (age$<5$~Gyr), metal-rich stars in the bulge region ($r_{\rm guide}<3$~kpc) has been confirmed with many independent observations \citep{bensby2013,buell2013,gesicki2014,schultheis2017,bernard_2018_bulgeSFH,hasselquist2020}. The number of young, metal-poor stars are even much lower (11 sub-solar metallicity bulge stars in the 2~Gyr age bin) and their presence need to be confirmed with further observations. In this work, for completeness, we include all of our young bulge stars in the analysis. 

Figures~\ref{mdf-decomp-age3} and \ref{mdf-decomp-age5} show the comparable best-fitted decompositions for the observed mono-age MDFs at ages of 3 and 5~Gyr, respectively. 
It is interesting to note that, compared to the mono-age population at 2~Gyr, the decomposition at age of 3~Gyr requires a larger contribution from stars born at other radii --- i.e., a larger radial migration effect. This is {qualitatively} consistent with the expectation that older stars have more time to migrate away from their birth radii, which is indeed seen in simulations \citep[e.g.,][]{halle2015,johnson2021}.

In contrast to the observed MDFs of 2 and 3~Gyr old stars, the MDFs of the 5~Gyr populations (Figure~\ref{mdf-decomp-age5}) are not well reproduced by the best-fitted simulated MDFs, particularly at intermediate radial bins (e.g., $7<r_{\rm guide}<9$~kpc and $9<r_{\rm guide}<11$~kpc). The simulated MDFs at these radial bins exhibit a clear density valley at $\rm [Fe/H] \sim -0.3$, while the observed MDFs show broad, monomodal distributions with no sign of density dip. This mismatch suggests that 
at least one of our assumptions described above does not hold at lookback time of 5~Gyr. One likely possibility is that the {birth} mono-age MDF at age of 5~Gyr at intermediate radius is not a single Gaussian profile as narrow as the outer disc but a broadened distribution that covers a wide range of [Fe/H]. Such a broad distribution in [Fe/H] is consistent with the predictions of a late accretion scenario proposed to explain the more complex age-chemistry structure of the disc, when other dimensions are considered \citep[e.g.,][]{lian2020a,lian2020b}. In this scenario, the Galactic disc experienced a recent significant gas accretion event $\sim6$~Gyr ago that rapidly diluted the abundances in the interstellar medium (ISM) from supersolar to subsolar values, on a short timescale of order 1~Gyr. During this period, many stars were formed with a narrow range in age but a wide range in [Fe/H], resulting in a complex mono-age {birth} MDF, instead of the simple Gaussian profiles assumed above.   
With current typical age uncertainties (e.g., $\delta_{\rm age}\sim2$~Gyr at $\rm age = 6$~Gyr), the observed sample at quoted $\rm age \sim 4-8$~Gyr would contain a notable fraction of these stars with an intrinsically complex MDF.  
Therefore, we focus below on the results of the decompositions at ages of 2 and 3~Gyr. To expand the decomposition to the stars at age around $\sim6$ Gyr, we would need more precise stellar ages (with uncertainties much smaller than 1~Gyr) to sufficiently resolve stellar populations formed at different stages of the accretion event.

\subsubsection{Radial migration strength}
\begin{figure*}
	\centering
	\includegraphics[width=18cm,viewport=10 10 950 430,clip]{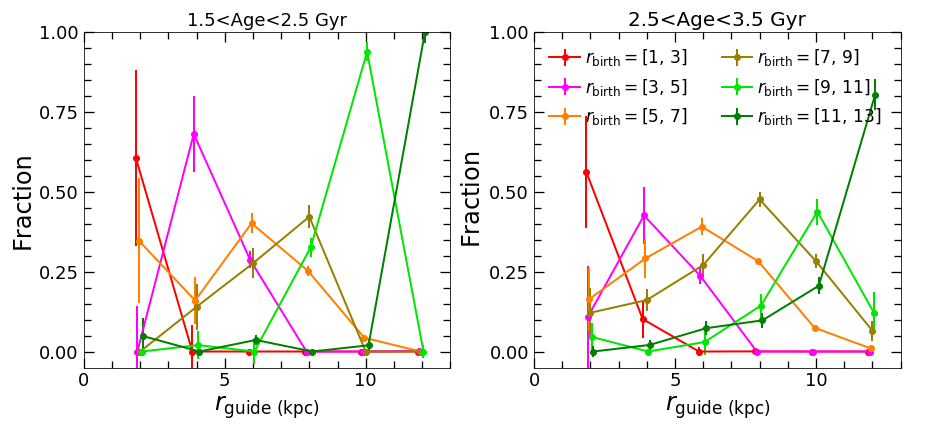}
	\caption{The contributed fraction of stars born at different birth radii (indicated by the colour of the lines) to bins of present-day guiding radii, derived from the decomposition of mono-age MDFs described in \textsection3.2.2. 
	The left panel shows the results for the stars of $\rm age = 2$~Gyr, and right panel for $\rm age = 3$~Gyr. 
	} 
	\label{migra-frac-rguide}
\end{figure*} 

\begin{figure*}
	\centering
	\includegraphics[width=18cm,viewport=10 10 950 430,clip]{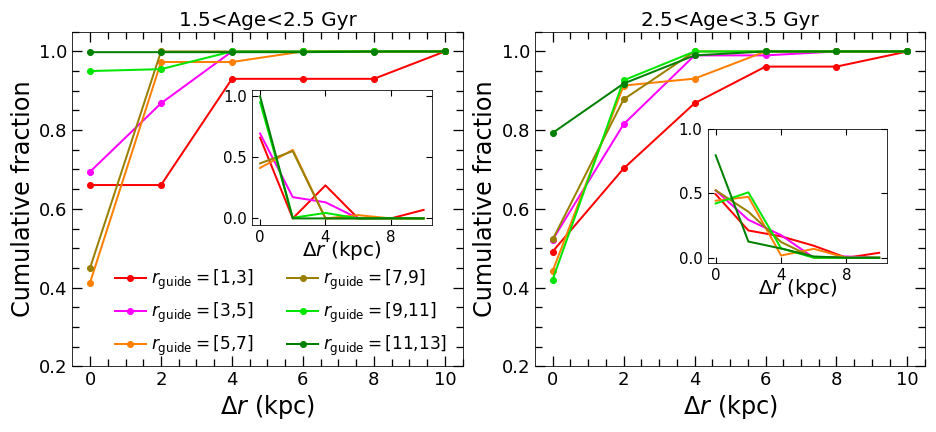}
	\caption{Cumulative fraction of stars formed at different distances from their current radii. Each line indicates one present-day radial bin as illustrated in the legend in the left panel. The first point on each line indicates the fraction of stars formed locally. Data points at larger $\Delta r$ include stars that migrated from larger distances to the present radius. {Inserted panel: contributed fraction at each present radius and age bin as a function of migration distance. The colour scheme is the same as the main plot.} 
	} 
	\label{migra-frac-rbirth}
\end{figure*} 

{Figure~\ref{migra-frac-rguide} shows the decomposed fraction (i.e. $\alpha_{i}$) from different birth radii at each present-day radius (left panel: 2~Gyr stars; right panel: 3~Gyr stars).} 
The decomposition results are also listed in Table~\ref{decomp}. 
The y-axis of each plot indicates the fraction of stars with a given current $r_{\rm guide}$ that were ``contributed'' from each $r_{\rm birth}$ bin, indicated by the colour of the lines.
This contributed fraction always peaks at the local radial bin (i.e., $r_{\rm guide} = r_{\rm birth}$) and drops rapidly in- and outward. This result suggests a generally minor radial migration effect of young stars, with decreasing impact from migration over larger distances. 

\begin{table*}
    \caption{Contributed fraction of stars born at different birth radii to each present radius (shown in Figure~\ref{migra-frac-rguide}).}
    \centering
    \begin{tabular}{ c c c c c c c c}
    \hline\hline 
    Age & Birth radius & \multicolumn{6}{c}{Present radius} \\
    \cmidrule(lr){3-8}
    &  & $[1,3]$ & $[3,5]$ & $[5,7]$ & $[7,9]$ & $[9,11]$ & $[11,13]$ \\
    \hline
 & [1 ,3] & 0.605$\pm$0.276 & 0.0$\pm$0.083 & 0.0$\pm$0.0 & 0.0$\pm$0.0 & 0.0$\pm$0.0 & 0.0$\pm$0.001\\
 & [3 ,5] & 0.0$\pm$0.144 & 0.681$\pm$0.12 & 0.287$\pm$0.027 & 0.0$\pm$0.003 & 0.0$\pm$0.001 & 0.002$\pm$0.001\\
 2~Gyr & [5 ,7] & 0.347$\pm$0.195 & 0.16$\pm$0.073 & 0.401$\pm$0.031 & 0.253$\pm$0.016 & 0.043$\pm$0.007 & 0.0$\pm$0.001\\
 & [7 ,9] & 0.0$\pm$0.006 & 0.14$\pm$0.072 & 0.276$\pm$0.047 & 0.422$\pm$0.036 & 0.0$\pm$0.004 & 0.0$\pm$0.0\\
 & [9 ,11] & 0.0$\pm$0.0 & 0.02$\pm$0.044 & 0.0$\pm$0.04 & 0.326$\pm$0.03 & 0.938$\pm$0.03 & 0.0$\pm$0.021\\
 & [11 ,13] & 0.048$\pm$0.056 & 0.0$\pm$0.001 & 0.036$\pm$0.017 & 0.0$\pm$0.0 & 0.019$\pm$0.022 & 0.998$\pm$0.034\\
 \hline
 & [1 ,3] & 0.562$\pm$0.175 & 0.102$\pm$0.059 & 0.0$\pm$0.014 & 0.0$\pm$0.002 & 0.0$\pm$0.0 & 0.0$\pm$0.0\\
 & [3 ,5] & 0.108$\pm$0.159 & 0.426$\pm$0.09 & 0.236$\pm$0.027 & 0.0$\pm$0.0 & 0.0$\pm$0.0 & 0.0$\pm$0.0\\
 3~Gyr & [5 ,7] & 0.163$\pm$0.098 & 0.291$\pm$0.06 & 0.391$\pm$0.027 & 0.283$\pm$0.011 & 0.074$\pm$0.007 & 0.01$\pm$0.006\\
 & [7 ,9] & 0.121$\pm$0.079 & 0.161$\pm$0.035 & 0.271$\pm$0.034 & 0.476$\pm$0.025 & 0.283$\pm$0.021 & 0.065$\pm$0.031\\
 & [9 ,11] & 0.046$\pm$0.044 & 0.0$\pm$0.008 & 0.03$\pm$0.041 & 0.144$\pm$0.037 & 0.437$\pm$0.041 & 0.122$\pm$0.063\\
 & [11 ,13] & 0.0$\pm$0.018 & 0.021$\pm$0.017 & 0.073$\pm$0.021 & 0.097$\pm$0.023 & 0.206$\pm$0.026 & 0.803$\pm$0.048\\
    \hline
    \end{tabular}
    \label{decomp}
\end{table*}

Figure~\ref{migra-frac-rbirth} expands on Figure~\ref{migra-frac-rguide} to show 
the cumulative fraction of stars that have moved some distance $\Delta r$ from their birth radius to their present-day $r_{\rm guide}$, indicated by lines of the same colors used in Figures~\ref{mdf-decomp-age2}--\ref{mdf-decomp-age5}.
That is, the first point ($\Delta r = 0$~kpc) indicates the fraction of stars formed locally at each present-day $r_{\rm guide}$, and the second point ($\Delta r = 2$~kpc) denotes the fraction of stars formed {\it up to} one of our radial bins away, and so on. {The inserted panel presents the contributed fraction at each present radius and age bin as a function of migration distance.}
It is clearly seen here that radial migration only contributes a minor fraction of the young stars at any present radius. At an age of 2~Gyr (left panel), the local contributions at most present radii are higher than 60\%, i.e., more than half of the 2~Gyr stars we observe today were formed {locally}. This fraction even goes {above} 95\% for the bin at $11<r_{\rm guide}<13$~kpc (80\% for 3~Gyr-old stars). 
These fractions reach $>$95\% ($>$90\%) after including stars that have migrated less than 4~kpc in 2~Gyr (3~Gyr), suggesting highly inefficient long distance radial migration. 

Interestingly, the innermost radial bin shows a systematically lower contribution from local and nearby components, indicating relatively more contribution from stars migrated from larger birth radii. This is broadly consistent with the inactive local star formation in the bulge region in the recent Universe \citep{zoccali2003,grieco2012,nataf2016,hasselquist2020,lian2020c}. However, the lower fraction of local and nearby contribution at age of 2~Gyr compared to the 3~Gy age bin is opposite to the trend seen in other radial bins. This is likely due to the noisy mono-age MDF of innermost radial bin at age of 2~Gyr that shows an unconfirmed excess of metal-poor stars as discussed in \textsection~4.1.2.  

To summarize, at young ages ($<$3~Gyr), {comparing to migrated populations, a large fraction ($>80$\%)} of disc stars we observe today were formed within 2~kpc of their present radius, and around half of them were formed locally (within 1~kpc). Very few young stars have migrated more than 4~kpc away from their birth radii. 
These results suggest that radial migration played a limited role in the recent evolution history of the Galactic disc {comparing to the local star formation}.

\subsection{Origin of the very outer stellar disc beyond $r>13$~kpc}
\begin{figure*}
	\centering
	\includegraphics[width=18cm,viewport=10 10 1470 860,clip]{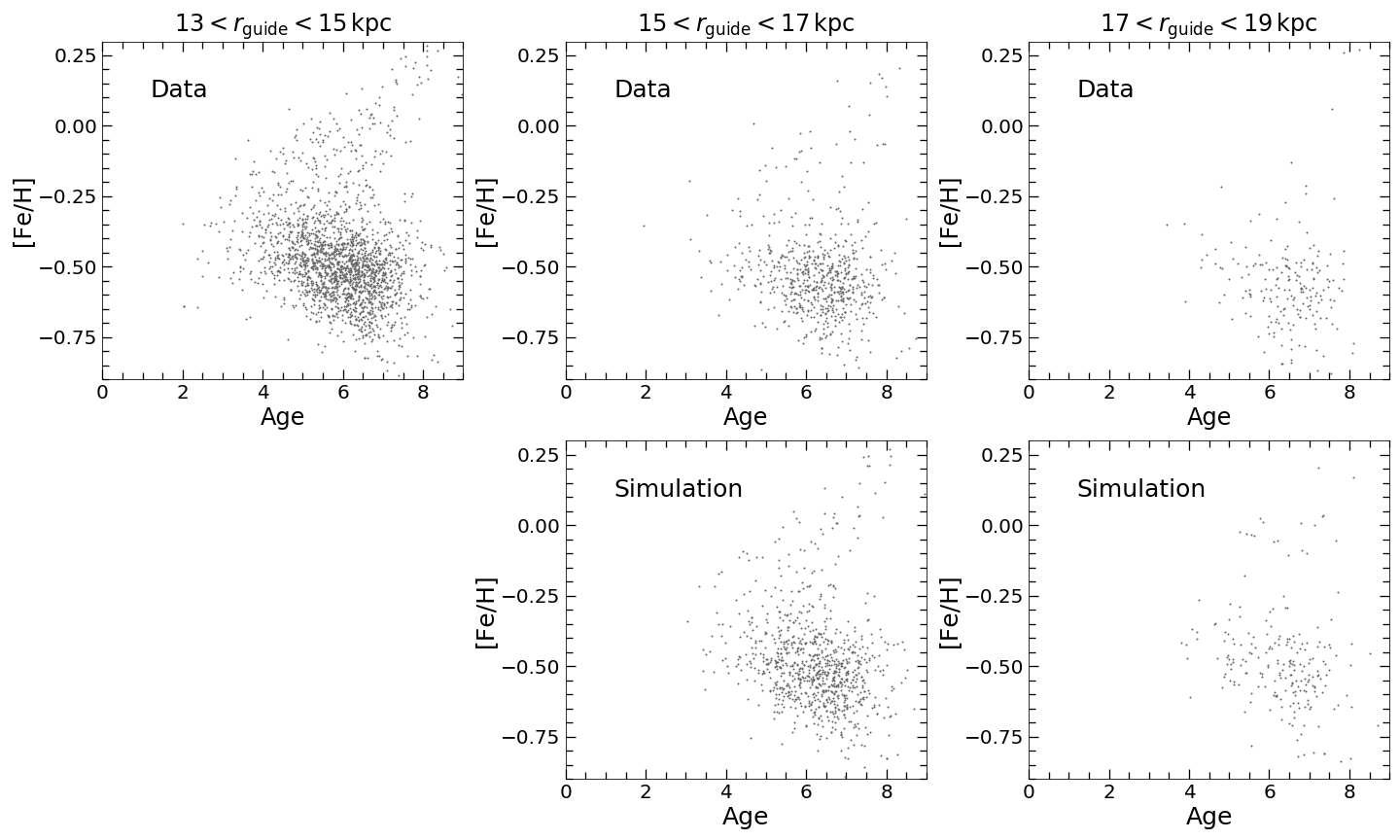}
	\caption{Comparison between the observed (top row) and simulated distributions (bottom row) in the age--[Fe/H] plane at different radii. The simulation starts with the distribution at $13<r_{\rm guide}<15$~kpc, and then predicts the distribution at larger radial bins considering only radial migration processes. 
	} 
	\label{migra-simulation}
\end{figure*} 

\begin{figure*}
	\centering
	\includegraphics[width=18cm,viewport=10 10 1470 860,clip]{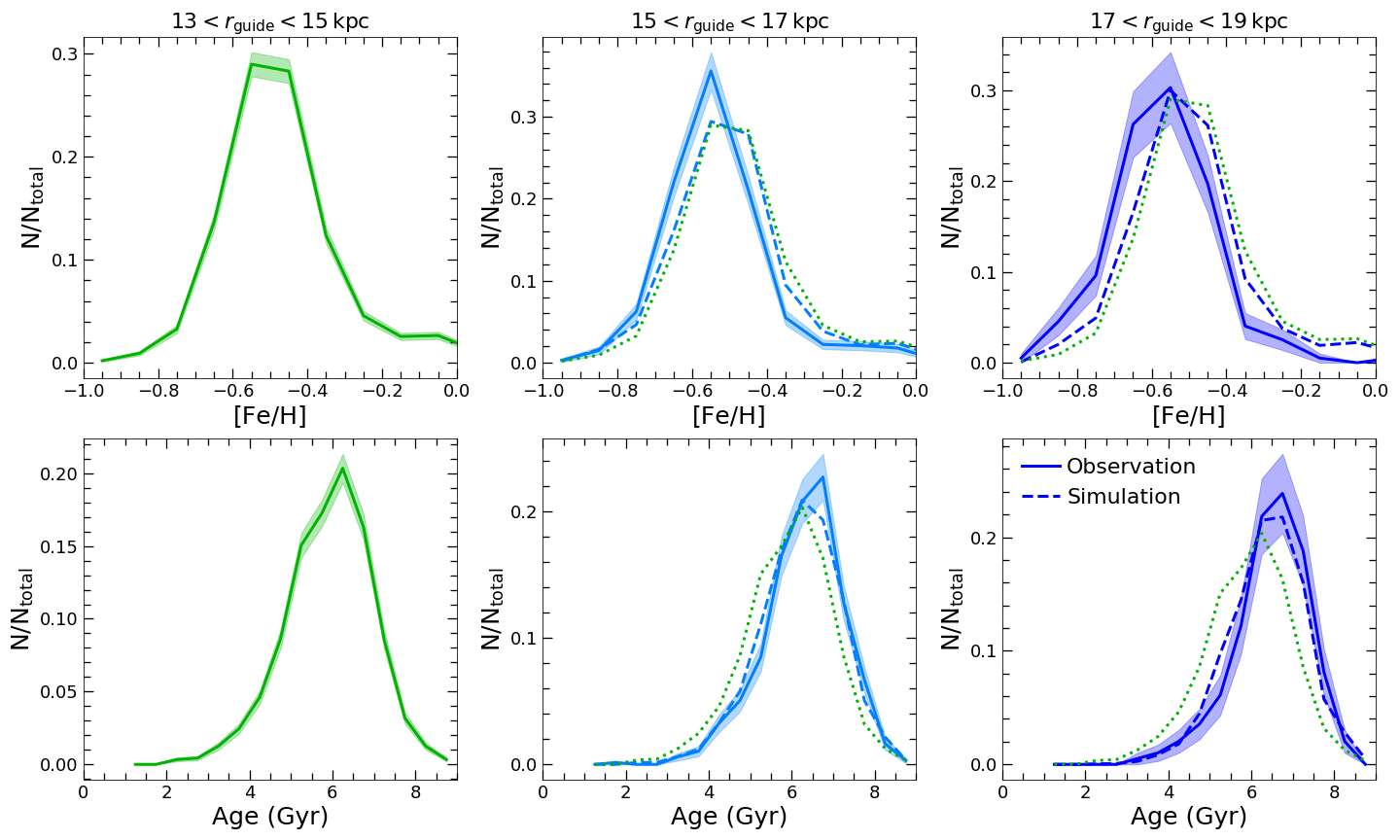}
	\caption{Comparison between the observed and simulated distributions in [Fe/H] (top row) and age (bottom row). Each column indicates one radial bin. The solid line in each panel denotes the observed [Fe/H] or age distribution, and shaded region represent Poisson errors. In the two outer radial bins, {the dotted line denotes the observed distribution in the donor radial bin for reference and} the {dashed} {line represents} the simulated distribution. 
	} 
	\label{migra-mdf-adf}
\end{figure*} 

In \textsection4.1 we analyzed the age--[Fe/H] distribution at $r_{\rm guide}<13$~kpc. Here we focus on the {very} outer disc regions at $r_{\rm guide}>13$~kpc. As no artificial break is seen in the age--[Fe/H] distribution of outer disc stars (\textsection2), in this section we consider the full range of [Fe/H] of the low-$\alpha$ stars in the {very} outer disc.
{As mentioned in \textsection 2} that the ages of low-$\alpha$ stars with ${\rm [Fe/H] <-0.5}$ are determined by extrapolation from more metal-rich stars, and thus we consider the results in this section preliminary but nevertheless interesting to explore.  

As discussed in \textsection3.1, the {very} outer disc {beyond 13~kpc} presents a very different radial variation pattern in the age--[Fe/H] plane than the majority of the disc, with a higher concentration of older, lower-metallicity stars along the {young} age--[Fe/H] sequence at increasingly larger radii (Figure~\ref{age-feh-radius}). {This implies that the Milky Way galaxy also presents a positive age gradient in the outer skirt that has been observed in many other disc galaxies \citep[e.g.,][]{roskar2008b,ruiz-lara2017}.} 
This pattern is qualitatively consistent with the prediction of a scenario in which the {very} outer disc is populated via radial migration from inner regions. Older stars, born before those inner regions had enriched to their present-day values, have more time to migrate and therefore can reach larger distances. In addition to radial migration, a finely-tuned ``outside-in'' quenching scenario \citep{schaefer2017,lin2019}, with an earlier cessation of star formation at larger radii, could also potentially be able to explain the change in shape of the age--[Fe/H] distribution at $r_{\rm guide}>13$~kpc. {It is worth noting that this radial pattern of age--[Fe/H] distribution persists, although being less significant, if restricted to the log($g$) range of stars in the outermost bin (0$<{\rm log}(g)<2$). To minimize possible contribution from selection effect which may alter the log($g$) distribution, in the following analysis we focus on very outer disc stars with 0$<{\rm log}(g)<2$.}

In this paper, we explore the radial migration scenario by simulating the age--[Fe/H] distribution in the {very} outer disc as predicted by radial migration and comparing with the observations. We choose the radial bin at $13<r_{\rm guide}<15$~kpc as the innermost bin that contributes stars to larger radii, and we simulate the resulting age--[Fe/H] distribution beyond this radius, considering purely radial migration effects (i.e., zero {local} star formation). This choice of starting radius takes advantage of its broad span in age distribution and minimal contamination of stars not on the {young} age--[Fe/H] sequence. 

At any given time, we assume that a certain fraction, $\lambda$, of stars formed in the donor radial bin migrate to its adjacent larger radial bin. This translates into the differential equation for the donor radial bin:
\begin{equation}
    \frac{dN_1}{dt} = - \lambda N_1(t),
\end{equation}
where $N_1(t)$ is the {observed} number of stars of age $t$ in the donor radial bin. The solution of this equation is
\begin{equation}
    N_1(t) = N_{0} e^{- \lambda t}.
\end{equation}
{Here $N_0$ represents the number of stars of age $t$ born in the donor radial bin.
For the intermediate radial bin, there are stars both coming from the donor radial bin and leaving to the outer radial bin. Assuming the fraction $\lambda$ independent of radius and zero local star formation, we can write the differential equation for the intermediate radial bin as
\begin{equation}
    \frac{dN_2}{dt} = \lambda N_1(t) - \lambda N_2(t),
\end{equation}
where $N_2(t)$ is the number of stars in the intermediate radial bin. The first term in the right part of the equation describes the stars moving from the donor radial bin and the second represents those migrating to the outer radial bin. 
The solution of Equation~3 is
\begin{equation}
    N_2(t) = \lambda N_{0} t e^{- \lambda t}.
\end{equation}
Finally, for the outer radial bin, we only consider stars moving from intermediate radial bin without leaving to larger radii and therefore the differential equation can be written as
\begin{equation}
    \frac{dN_3}{dt} = \lambda N_2(t),
\end{equation}
with solution of 
\begin{equation}
    N_3(t) = N_0-N_0e^{-\lambda t}-\lambda N_{0} t e^{- \lambda t}.
\end{equation}}
Figure~\ref{migra-simulation} shows the results {of our simulation with migration {fraction} of {6\% per Gyr}} (bottom panels) and the comparison with observations (top panels) in the age--[Fe/H] plane. 
The simulation {well reproduces} the increasing concentration of stars towards older age with increasing radius. {Note that the normalized age-[Fe/H] distribution of simulation is not sensitive to the absolute value of the migration fraction}. 

Figure~\ref{migra-mdf-adf} compares the simulated and observed distributions from Figure~\ref{migra-simulation} separately in [Fe/H] (top row) and age (bottom row). The observed distributions are indicated with solid lines, while the simulated distributions are shown as {dashed} lines. 
It can be seen that the simulated distribution {well matches} the observed {age} distributions at $15<r_{\rm guide}<19$~kpc {{with} continuous shift towards older age at larger radii} {(i.e., positive age gradient)}. 
{The predicted [Fe/H] distribution shifts slightly towards lower [Fe/H] at larger radii, qualitatively consistent with the observed trend but quantitatively not sufficient to explain the data. This discrepancy is possibly related to the relatively more uncertain age measurements for metal-poor stars (\textsection{2}). A more significant shift in [Fe/H] distribution is expected if the ages of the metal-poor stars and therefore the number of these stars migrated to larger radii are underestimated.}


{Note that for simplicity we assume migration only comes from a single radial bin at $13<r_{\rm guide}<15$~kpc. In reality there will also have migrated stars from $<13$~kpc, which need to be considered when to quantify the migration strength in the very outer disc. To achieve that, a numerical simulation that considers the density distribution over a wide radial range is probably needed. Since the goal of this section is to qualitatively illustrate that the observed age-[Fe/H] distribution in the very outer disc at $r>15$~kpc can be explained with radial migration, we do not explore the more realistic and complex model in this work.}


\section{Discussion}
\subsection{Potential observational bias}
The results of this work rely on the radial comparison of stellar distributions in the age--[Fe/H] plane, which may be affected by potential bias in the sample selections at various radii. Here we discuss two potential sources of bias in the sample selection: the stellar parameters and the definition of radius. 

It has been shown that radial variations in APOGEE's stellar parameter distributions could introduce artificial variations in the stellar chemical abundance distributions (due to the survey's observational design; e.g., \citealt{griffith2020}). To verify our results against this potential bias, we resample the log($g$)--$T_{\rm eff}$ distribution at all radii to match that of the radial bin at $3<r_{\rm guide}<5$~kpc. 
Figure~\ref{logg-check} shows the median trend of the {median age-[Fe/H] relations} at different radii, after this resampling. The radial variation of these two sequences are consistent with that of the original sample shown in Fig.~\ref{age-feh-median} --- i.e., little radial dependence {at age$>$6~Gyr within solar radius and significant radial variation at age$<6$~Gyr}. 
We have also conducted mono-age MDF decompositions for the sample after resampling in log($g$)--$T_{\rm eff}$ space and found the fraction of {locally formed} and migrated stars to remain in line with the results presented in \textsection{4.1}. 
This suggests that our results are robust against the dependence of chemical abundances on stellar parameters. 

We also inspect the dependence of our results on our choice of Galactic radius. The guiding radius of a star represents the circular orbit that has the same angular momentum of the star, which could be significantly different from its instantaneous Galactocentric radius when the star's orbit has non-zero eccentricity and/or strong epicycle motion. The determination of a star's guiding radius relies on the knowledge of its position in the Galaxy and the Galactic gravitational potential well. 
Thus the measurement of a star's present location (e.g., Galactocentric radius) is more direct and precise compared to the measurement of the star's average location (e.g., the guiding radius). To verify whether the radial trend reported here is dependent on the choice of the radius definition, we select our sample using Galactocentric radius and show the comparison of the age--[Fe/H] sequences at various radii in Figure~\ref{radius-check}. The radial variations of the two age-[Fe/H] sequences are remarkably consistent with the sample selected by guiding radius shown in Fig.~\ref{age-feh-median}. A similar mono-age MDF analysis (\textsection{4.1}) is also performed for this sample selected by Galactocentric radius. The obtained strength of radial migration is in good consistency with our previous results. 
Therefore our results are not significantly dependent on the usage of either guiding or Galactocentric radius. 

\subsection{Comparison to other studies}
\begin{table}
    \caption{Migration distance estimates from simulations or observations in previous works. All estimates are in Galactocentric radius. The migration distance distribution is assumed to follow a half normal distribution to convert different migration distance estimates to the same basis and enable direct comparison between them.}
    \centering
    \begin{tabular}{ l c c}
    \hline\hline 
    Reference & Mean (Std) & Comment \\
    & migration distance & \\
     & kpc & \\
    \hline
 \citet{roskar2008a} & 2.4 (3.0) $^a$ & Simulation \\  
 \citet{kubryk2015} & 2.8 (3.6) $^b$ & Simulation \\
 \citet{halle2015} & 1.4 (1.8) $^b$  & Simulation\\
 \citet{frankel2018} & 1.7 (2.2) $^b$ & Observation\\
 \citet{frankel2020} & 1.4 (1.8) $^b$ & Observation \\
 \citet{silva2021} & 2.1 (2.7) $^a$ & Simulation \\
 \citet{khoperskov2021} & 2.4 (3.0) $^a$ & Simulation \\
    \hline
    \end{tabular}
    \label{migration-distance-lite}
    Note $^a$: migration distance over 10~Gyr. \\
         $^b$: migration distance over 3~Gyr. 
\end{table}

\begin{table}
    \caption{Mean (standard deviation of) migration distance at each present radius and age bin.}
    \centering
    \begin{tabular}{ l c c}
    \hline\hline 
    Age & Present radius & Migration distance \\
    Gyr & kpc & kpc \\
    \hline
 & 1$<r_{\rm guide}<$3 & 2.17 (2.71)\\
 & 3$<r_{\rm guide}<$5 & 1.34 (1.67)\\
2 & 5$<r_{\rm guide}<$7 & 1.54 (1.92)\\
 & 7$<r_{\rm guide}<$9 & 1.37 (1.71)\\
 & 9$<r_{\rm guide}<$11 & 0.68 (0.85)\\
 & 11$<r_{\rm guide}<$13 & 0.51 (0.64)\\
 \hline
 & 1$<r_{\rm guide}<$3 & 2.24 (2.80)\\
 & 3$<r_{\rm guide}<$5 & 1.81 (2.25)\\
3 & 5$<r_{\rm guide}<$7 & 1.76 (2.20)\\
 & 7$<r_{\rm guide}<$9 & 1.48 (1.85)\\
 & 9$<r_{\rm guide}<$11 & 1.49 (1.86)\\
 & 11$<r_{\rm guide}<$13 & 0.96 (1.20)\\
    \hline
    \end{tabular}
    \label{migration-distance}
\end{table}

While widely regarded as an important process shaping the disc structure within the Milky Way and other galaxies, {constraining the strength of radial migration from observations is difficult.}
{Current measurements of radial migration strength are mostly made in simulated galaxies. }
One of the first detailed quantifications of radial migration strength in numerical simulations was conducted by \citet{halle2015}. By examining an $N$-body simulation of an Sb-type disc galaxy, \citet{halle2015} estimated the migration fraction (number of migrators compared to the entire population) as a function of both radius and time. In this bar-dominated simulated galaxy, the migration strength peaks at radius of the bar cororation resonance. 
Outward migration is generally more efficient than inward migration, except at the innermost region, owing to the negative density profile, {and the} outward migration distance increases with birth radius. {Within 3~Gyr, 68\% stars have migrated by 1.8~kpc in Galactocentric radius. Only} a small fraction ($\sim$5\%) of stars moved farther than 4~kpc away {in terms of guiding radius} from their birth {places}. {Over an longer evolution time of 9~Gyr, the migration distance increases to 2.9~kpc and fraction of long distance migration ($>4$~kpc) is less than 20\%.} These results are in good consistency with our estimates obtained from observations described in \textsection4.1. 

A {comparable strength of} radial migration was reported in a Milky Way-like simulation by \citet{roskar2008a} (see also \citealt{loebman2011}). In this simulation, while 25\% of the entire disc stellar population migrated more than 2~kpc over an evolution time of 10~Gyr, more than half of the stars in the solar cylinder today (7.5$<r_{\rm GC}<$8.5~kpc) moved at least 2~kpc from their birth radius \citep{roskar2008b}. {Assuming a Gaussian profile for the migration distance distribution, this suggests an average migration distance of 2.4~kpc and 68\% stars migrated by up to 3.0~kpc.} A similar strength of radial migration is {also found in other simulations} \citep{silva2021,khoperskov2021}. 
{\citet{silva2021} reported a fraction of 55\% and 35\% in the old thin and thick disc stars migrated from their birth radii by more than 2~kpc.}
{A stronger radial migration was found by \citet{kubryk2013} in a simulated disk galaxy with a strong and long bar. Adopting an empirical correction to account for the difference in bar strength between this simulated galaxy and the Milky Way, the average migration distance of stars born at solar radius and 3~Gyr ago is 2.8~kpc \citep{kubryk2015}. Interestingly, the radial migration distance in this simulated galaxy decreases with birth radius and time, which are qualitatively consistent with the findings in our work.}
{\citet{khoperskov2021} found a fraction of 50\% stars in their simulated galaxies move up to 2~kpc from their birth radii and about 10\% stars are extreme migrators with migration distance greater than 5~kpc. Interestingly, they found that the bimodal [$\alpha$/Fe]-[Fe/H] distribution is mainly established by the local star formation and chemical enrichment process during the thick and thin disc formation and the global chemical pattern is not strongly reshaped by the radial migration (see a different result in \citet{sharma2020}).} 
{Using $N$-body simulations, \citet{minchev2011} found that a strong radial migration in churning mode (i.e. efficient exchange of angular momentum) can be induced by resonance overlap of multiple patterns (e.g., bars and spiral arms).} 

{To enable a rough but direct comparison between the radial migration strength measured in different works, we convert the various references to migration strength to the same definition, i.e. mean migration distance of the whole population. A Gaussian migration distance distribution is assumed for this conversion. For reference, we also calculate the standard deviation of migration distance distribution, which is the distance that 68\% stars have stayed within their birth radii. 
Table~\ref{migration-distance-lite} includes the converted mean and standard deviation of migration distance reported in the simulation and observation works mentioned above. Note that all estimates are in Galactocentric radius and in some simulations only the migration distance over a long evolution time of 10~Gyr is available.
Interestingly, after considering the different timescale of these estimates, the strength of radial migration measured in different simulated galaxies, except for that in \citet{kubryk2013}, are roughly consistent.} 

Unlike the direct measurements from simulation, quantification of radial migration strength from observations is much more difficult, due to the complex interplay of multiple astrophysical processes (e.g., gas accretion, star formation, and radial migration) and observational uncertainties and systematics that all shape the observations. Thanks to the advancements in observations in the last decade, we are now able to obtain robust stellar chemistry and age information across a large portion of the Galaxy, which greatly improves our ability to infer properties of radial migration from the data. 

Based on an empirical model that allows radial migration to vary, \citet{frankel2018} quantitatively fit the observed age--[Fe/H] distribution of red clump stars (from the APOGEE survey) as a function of radius within $5<r_{\rm GC}<15$~kpc. The best-fitted radial migration {parameter suggests 68\% stars have migrated within a} distance of 3.6$\sqrt{\tau/8{\rm Gyr}}$~kpc, where $\tau$ indicates the length of the evolution time. {The migration distance was assumed to follow a Gaussian distribution.} {It is implied that} stars at age of 3~Gyr have moved systematically, either in- or outward, by {1.7}~kpc on average ({or up to 2.2~kpc for 68\% stars}). A {lower value of average migration distance (1.4~kpc) was suggested} by \citet{frankel2020} using more recent data (APOGEE DR14) and updated models with more complex form of metallicity radial profile. 

{At} an age of 3~Gyr, we find half of the stars at most {present} radii were formed locally, with average radial displacement less than 1~kpc, and another one-third formed nearby with an average displacement of 2~kpc. {If assuming an average displacement of 0.5~kpc for the stars formed in the local radial bin, we calculate mean and standard deviation of migration distance for each present radius and age bin, which are listed in Table~\ref{migration-distance}. Except for the inner most radial bin, the average migration distance is 0.5$-$1.6~kpc at age of 2~Gyr and 1.0$-$1.8~kpc at age of 3~Gyr. {These results are in good agreement with the estimate in the Milky Way in \citet{frankel2020} as well as in simulated galaxies as summarized in Table~3.} 
{Our estimates of} migration distance at the inner disc within solar radius {are} higher than that in the disc beyond, suggesting that the bar might be more effective than the spiral arm in driving radial migration.} 


 

\subsection{Radial migration's effect on the Galactic metallicity gradient}
\begin{figure}
	\centering
	\includegraphics[width=16cm,viewport=10 10 1100 500,clip]{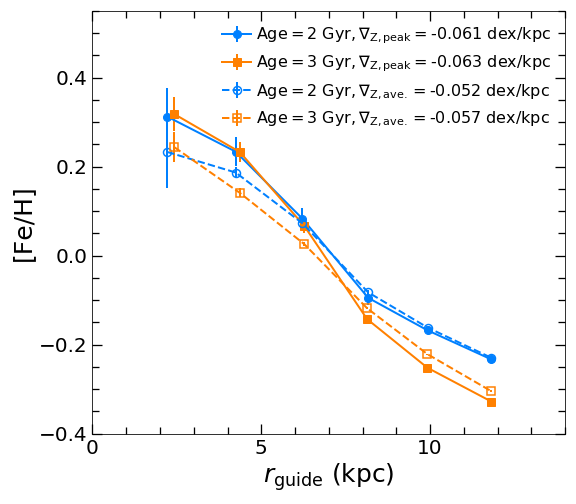}
	\caption{Radial values of the intrinsic [Fe/H] (filled symbols, solid lines) and average [Fe/H] (open symbols, dashed lines) for mono-age populations at 2~Gyr (blue circles) and 3~Gyr (orange squares).    
	} 
	\label{Z-gradient}
\end{figure} 

The assumption adopted here that the peak [Fe/H] of the observed MDF is the same as the {birth} MDF allows us to map the radial distribution of the [Fe/H] of the ISM at a given look-back time. From the observed MDF we derive the mean [Fe/H], whose radial distribution incorporates the effect of radial migration. Figure~\ref{Z-gradient} shows the radial distribution of the intrinsic and average [Fe/H] for mono-age populations at 2 and 3~Gyr. We calculate the radial gradients of the intrinsic and present average [Fe/H] for each mono-age population, which are shown in the top-right legend. It is interesting to note that the metallicity gradient seems to be steepest at intermediate radial bins and flattens at both smaller and larger radii. 

The intrinsic metallicity gradient at an age of 2~Gyr is $-$0.061~dex/kpc, which is comparable or slightly flatter than the metallicity gradient of young stars reported in \citet{bergemann2014} ($-$0.076~dex/kpc for stars with ages $<$7~Gyr and $|{\rm z}|$ within 300~pc), \citet{anders2017} ($-$0.066~dex/kpc at age between 1 and 2~Gyr), OB stars in \citet{braganca2019} ($-$0.07$-$0.09 dex/kpc at 8$<r_{\rm GC}<$16~kpc), \citet{hasselquist2019} ($-$0.06~dex/kpc), \citet{feuillet2019} ($-$0.059~dex/kpc), young open clusters in  \citet{zhang2021} ($-$0.074~dex/kpc), and of Cepheids in \citet{genovali2014} ($-$0.06~dex/kpc) and \citet{minniti2020} ($-0.062$~dex/kpc), but is steeper than that of H II regions derived by \citet{balser2011} ($\sim-$0.03-0.04~dex/kpc), OB stars in \citet{daflon2004} ($-$0.042~dex/kpc), and of young (age$<1$~Gyr) planetary nebulae in \citet{stanghellini2018} ($-$0.027~dex/kpc).  
For both the peak and average [Fe/H], the radial gradient is flatter at younger ages. Such flattening of metallicity gradient with time in the past few Gyrs is also reported in \citet{anders2017},\citet{minchev2018}, and \citet{hasselquist2019}. It is likely a result of a faster enrichment in the recent past (leading to a steeper slope of the age--[Fe/H] relation at young ages) in the outer disc, compared to the inner regions. Compared to the peak [Fe/H], the radial gradient of average [Fe/H] is flatter, a result of extra mixing caused by radial migration. Such flattening of the metallicity gradient due to radial migration is also predicted by Galaxy evolution models and simulations \citep[e.g.,][]{kubryk2013,minchev2014,vincenzo2020}. 


\section{Summary}
In this work we investigate observational constraints on the radial migration process using age--[Fe/H] distributions across the Galaxy. 
We find that the Milky Way's stars fall in two distinct age-[Fe/H] sequences: an early sequence that dominates the inner Galaxy and a late one most prevalent in the outer Galaxy. At intermediate solar radius, these age--[Fe/H] sequences overlap, resulting in a complex age--[Fe/H] pattern that implies a complex star formation history in the solar neighborhood. 
{By comparing the median age-[Fe/H] relation at different radii, we find that the disc at $r>6$~kpc presents a broken age-[Fe/H] relation with a more pronounced break at larger radii, which is possibly a result of a metal-poor gas accretion event. At a given age, the metallicity is systematically lower at larger radii at age $<6$~Gyr, but nearly constant with radius at age $>6$~Gyr.}
We have confirmed that these results are robust against the systematic dependence of abundances on stellar parameters and against our choice of guiding or Galactocentric radius.    

To obtain observational constraints on the radial migration and minimize the effect of a radially variant star formation history, we perform a detailed analysis of the MDFs of mono-age populations at different radii within 13~kpc of the Galactic center. From the outer disc to the inner Galaxy, the mono-age MDFs become significantly broader, with more pronounced tails at low [Fe/H]. This trend can be explained by radial migration effects. 
Given the radial variation of [Fe/H] at a given age for young stars, we use [Fe/H] as a tracer of a star's birth radius and decompose the mono-age MDFs at 2, 3, and 5~Gyr at each present-day radius into different Gaussian components originating from various birth radii. The reconstructed MDFs  match well the observed ones at 2 and 3~Gyr, but are inconsistent with the observations at 5~Gyr. The discrepancy at 5~Gyr implies that the {birth} MDF at this age may not be a narrow symmetric Gaussian profile but rather span a wide range in [Fe/H], likely due to a recent dilution process triggered by a late gas accretion event that occurred around $\sim6$~Gyr ago \citep{lian2020a}. The radial distribution of the peak [Fe/H] of the mono-age MDF is steeper than that of the average [Fe/H] for both mono-age populations at age of 2 and 3~Gyr, indicative of a flattening of the metallicity gradient due to radial migration. 

The decomposition results suggest that, for stars with age of 2 and 3~Gyr, {about} half of them were formed locally (within 1~kpc of their present radius) and the majority {(above 80\%)} formed within 2~kpc on average. Very few stars (fewer than 5\%) were formed farther than 4~kpc away from their present radius, suggesting inefficient long distance migration in the recent past of the Milky Way. {Assuming a migration distance of 0.5~kpc for stars remained in the birth radial bin, we obtain an average migration distance of 0.5$-$1.6 and 1.0$-1.8$~kpc at age of 2 and 3~Gyr, respectively, which are in good consistence with other estimates of radial migration strength measured in simulated galaxies and our Milky Way on the same timescale \citep{halle2015,frankel2020}.}

At radii beyond 13~kpc, there is a clear trend of increasing concentration of stars in the old, low-metallicity regime with increasing radius. This radial trend results in a negative metallicity gradient and a positive age gradient in the very outskirts of the Galactic disc. To test whether this radial trend can be explained by radial migration, we conduct a simple test that takes the observations at $13<r_{\rm guide}<15$~kpc and mimics the distribution in age--[Fe/H] at larger radii considering radial migration under the assumption that no stars are formed {locally} at those radii during this period. We find that the observed age--[Fe/H] distribution, {especially the age distribution,} beyond 15~kpc can be well explained by radial migration alone. 

The results presented in this paper impose strong constraints on the radial migration strength in the Milky Way, which could also be useful references for Milky Way-like galaxy simulations. 

\section*{Acknowledgements}
We are grateful to the referee for the constructive report that greatly improved the clarity and readability of the paper.
JL is grateful to Neige Frankel, Sarah Loebman and Nikos Prantzos for useful discussions and suggestions. This material is based upon work supported by the National Science Foundation under Grant No. 2009993. SH was supported by an NSF Astronomy and Astrophysics Postdoctoral Fellowship under award AST-1801940. DAGH acknowledges support from the State Research Agency (AEI) of the Spanish Ministry of Science, Innovation and Universities (MCIU) and the European Regional Development Fund (FEDER) under grant AYA2017-88254-P. DT acknowledges support from the Science, Technology
and Facilities Council through the Consolidated Grant Cosmology and Astrophysics at Portsmouth, ST/S000550/1. J.G.F-T gratefully acknowledges the grant support provided by Proyecto Fondecyt Iniciaci\'on No. 11220340, and also from ANID Concurso de Fomento a la Vinculaci\'on Internacional para Instituciones de Investigaci\'on Regionales (Modalidad corta duraci\'on) Proyecto No. FOVI210020, and from the Joint Committee ESO-Government of Chile 2021 (ORP 023/2021).

Funding for the Sloan Digital Sky Survey IV has been provided by the Alfred P. Sloan Foundation, the U.S. Department of Energy Office of Science, and the Participating Institutions. SDSS-IV acknowledges
support and resources from the Center for High-Performance Computing at
the University of Utah. The SDSS web site is www.sdss.org.

SDSS-IV is managed by the Astrophysical Research Consortium for the 
Participating Institutions of the SDSS Collaboration including the 
Brazilian Participation Group, the Carnegie Institution for Science, 
Carnegie Mellon University, the Chilean Participation Group, the French Participation Group, Harvard-Smithsonian Center for Astrophysics, 
Instituto de Astrof\'isica de Canarias, The Johns Hopkins University, Kavli Institute for the Physics and Mathematics of the Universe (IPMU) / 
University of Tokyo, the Korean Participation Group, Lawrence Berkeley National Laboratory, 
Leibniz Institut f\"ur Astrophysik Potsdam (AIP),  
Max-Planck-Institut f\"ur Astronomie (MPIA Heidelberg), 
Max-Planck-Institut f\"ur Astrophysik (MPA Garching), 
Max-Planck-Institut f\"ur Extraterrestrische Physik (MPE), 
National Astronomical Observatories of China, New Mexico State University, 
New York University, University of Notre Dame, 
Observat\'ario Nacional / MCTI, The Ohio State University, 
Pennsylvania State University, Shanghai Astronomical Observatory, 
United Kingdom Participation Group,
Universidad Nacional Aut\'onoma de M\'exico, University of Arizona, 
University of Colorado Boulder, University of Oxford, University of Portsmouth, 
University of Utah, University of Virginia, University of Washington, University of Wisconsin, 
Vanderbilt University, and Yale University.

\section*{Data Availability}
The data underlying this article is from an internal incremental release of the SDSS-IV/APOGEE survey, following the SDSS-IV public Data Release 16 (using reduction pipeline version r13). 
This incremental catalog is anticipated to be made public in a future post-DR17 release.

\bibliographystyle{mnras}
\bibliography{Jianhui}{}

\begin{thebibliography}{}
\makeatletter
\relax
\def\mn@urlcharsother{\let\do\@makeother \do\$\do\&\do\#\do\^\do\_\do\%\do\~}
\def\mn@doi{\begingroup\mn@urlcharsother \@ifnextchar [ {\mn@doi@}
  {\mn@doi@[]}}
\def\mn@doi@[#1]#2{\def\@tempa{#1}\ifx\@tempa\@empty \href
  {http://dx.doi.org/#2} {doi:#2}\else \href {http://dx.doi.org/#2} {#1}\fi
  \endgroup}
\def\mn@eprint#1#2{\mn@eprint@#1:#2::\@nil}
\def\mn@eprint@arXiv#1{\href {http://arxiv.org/abs/#1} {{\tt arXiv:#1}}}
\def\mn@eprint@dblp#1{\href {http://dblp.uni-trier.de/rec/bibtex/#1.xml}
  {dblp:#1}}
\def\mn@eprint@#1:#2:#3:#4\@nil{\def\@tempa {#1}\def\@tempb {#2}\def\@tempc
  {#3}\ifx \@tempc \@empty \let \@tempc \@tempb \let \@tempb \@tempa \fi \ifx
  \@tempb \@empty \def\@tempb {arXiv}\fi \@ifundefined
  {mn@eprint@\@tempb}{\@tempb:\@tempc}{\expandafter \expandafter \csname
  mn@eprint@\@tempb\endcsname \expandafter{\@tempc}}}

\bibitem[\protect\citeauthoryear{{Adibekyan}, {Santos}, {Sousa}  \&
  {Israelian}}{{Adibekyan} et~al.}{2011}]{adibekyan2011}
{Adibekyan} V.~Z.,  {Santos} N.~C.,  {Sousa} S.~G.,   {Israelian} G.,  2011,
  \mn@doi [\aap] {10.1051/0004-6361/201118240}, \href
  {https://ui.adsabs.harvard.edu/abs/2011A&A...535L..11A} {535, L11}

\bibitem[\protect\citeauthoryear{{Ahumada} et~al.,}{{Ahumada}
  et~al.}{2020}]{ahumada2020}
{Ahumada} R.,  et~al., 2020, \mn@doi [\apjs] {10.3847/1538-4365/ab929e}, \href
  {https://ui.adsabs.harvard.edu/abs/2020ApJS..249....3A} {249, 3}

\bibitem[\protect\citeauthoryear{{Anders} et~al.,}{{Anders}
  et~al.}{2017}]{anders2017}
{Anders} F.,  et~al., 2017, \mn@doi [\aap] {10.1051/0004-6361/201629363}, \href
  {https://ui.adsabs.harvard.edu/abs/2017A&A...600A..70A} {600, A70}

\bibitem[\protect\citeauthoryear{{Babusiaux} et~al.,}{{Babusiaux}
  et~al.}{2010}]{babusiaux2010}
{Babusiaux} C.,  et~al., 2010, \mn@doi [\aap] {10.1051/0004-6361/201014353},
  \href {https://ui.adsabs.harvard.edu/abs/2010A&A...519A..77B} {519, A77}

\bibitem[\protect\citeauthoryear{{Bakos}, {Trujillo}  \& {Pohlen}}{{Bakos}
  et~al.}{2008}]{bakos2008}
{Bakos} J.,  {Trujillo} I.,   {Pohlen} M.,  2008, \mn@doi [\apjl]
  {10.1086/591671}, \href
  {https://ui.adsabs.harvard.edu/abs/2008ApJ...683L.103B} {683, L103}

\bibitem[\protect\citeauthoryear{{Balser}, {Rood}, {Bania}  \&
  {Anderson}}{{Balser} et~al.}{2011}]{balser2011}
{Balser} D.~S.,  {Rood} R.~T.,  {Bania} T.~M.,   {Anderson} L.~D.,  2011,
  \mn@doi [\apj] {10.1088/0004-637X/738/1/27}, \href
  {https://ui.adsabs.harvard.edu/abs/2011ApJ...738...27B} {738, 27}

\bibitem[\protect\citeauthoryear{{Balser}, {Wenger}, {Anderson}  \&
  {Bania}}{{Balser} et~al.}{2015}]{balser2015}
{Balser} D.~S.,  {Wenger} T.~V.,  {Anderson} L.~D.,   {Bania} T.~M.,  2015,
  \mn@doi [\apj] {10.1088/0004-637X/806/2/199}, \href
  {https://ui.adsabs.harvard.edu/abs/2015ApJ...806..199B} {806, 199}

\bibitem[\protect\citeauthoryear{{Beaton} et~al.,}{{Beaton}
  et~al.}{2021}]{beaton2021}
{Beaton} R.~L.,  et~al., 2021, arXiv e-prints, \href
  {https://ui.adsabs.harvard.edu/abs/2021arXiv210811907B} {p. arXiv:2108.11907}

\bibitem[\protect\citeauthoryear{{Bensby} et~al.,}{{Bensby}
  et~al.}{2013}]{bensby2013}
{Bensby} T.,  et~al., 2013, \mn@doi [\aap] {10.1051/0004-6361/201220678}, \href
  {https://ui.adsabs.harvard.edu/abs/2013A&A...549A.147B} {549, A147}

\bibitem[\protect\citeauthoryear{{Beraldo e Silva}, {Debattista}, {Nidever},
  {Amarante}  \& {Garver}}{{Beraldo e Silva} et~al.}{2021}]{silva2021}
{Beraldo e Silva} L.,  {Debattista} V.~P.,  {Nidever} D.,  {Amarante} J. A.~S.,
    {Garver} B.,  2021, \mn@doi [\mnras] {10.1093/mnras/staa3966}, \href
  {https://ui.adsabs.harvard.edu/abs/2021MNRAS.502..260B} {502, 260}

\bibitem[\protect\citeauthoryear{{Bergemann} et~al.,}{{Bergemann}
  et~al.}{2014}]{bergemann2014}
{Bergemann} M.,  et~al., 2014, \mn@doi [\aap] {10.1051/0004-6361/201423456},
  \href {https://ui.adsabs.harvard.edu/abs/2014A&A...565A..89B} {565, A89}

\bibitem[\protect\citeauthoryear{{Bernard}, {Schultheis}, {Di Matteo}, {Hill},
  {Haywood}  \& {Calamida}}{{Bernard} et~al.}{2018}]{bernard_2018_bulgeSFH}
{Bernard} E.~J.,  {Schultheis} M.,  {Di Matteo} P.,  {Hill} V.,  {Haywood} M.,
   {Calamida} A.,  2018, \mn@doi [\mnras] {10.1093/mnras/sty902}, \href
  {https://ui.adsabs.harvard.edu/abs/2018MNRAS.477.3507B} {477, 3507}

\bibitem[\protect\citeauthoryear{{Bland-Hawthorn} \&
  {Gerhard}}{{Bland-Hawthorn} \& {Gerhard}}{2016}]{bland2016}
{Bland-Hawthorn} J.,  {Gerhard} O.,  2016, \mn@doi [\araa]
  {10.1146/annurev-astro-081915-023441}, \href
  {https://ui.adsabs.harvard.edu/abs/2016ARA&A..54..529B} {54, 529}

\bibitem[\protect\citeauthoryear{{Blanton} et~al.,}{{Blanton}
  et~al.}{2017}]{blanton2017}
{Blanton} M.~R.,  et~al., 2017, \mn@doi [\aj] {10.3847/1538-3881/aa7567}, \href
  {https://ui.adsabs.harvard.edu/abs/2017AJ....154...28B} {154, 28}

\bibitem[\protect\citeauthoryear{{Bonanno}, {Schlattl}  \&
  {Patern{\`o}}}{{Bonanno} et~al.}{2002}]{bonanno2002}
{Bonanno} A.,  {Schlattl} H.,   {Patern{\`o}} L.,  2002, \mn@doi [\aap]
  {10.1051/0004-6361:20020749}, \href
  {https://ui.adsabs.harvard.edu/abs/2002A&A...390.1115B} {390, 1115}

\bibitem[\protect\citeauthoryear{{Bovy}}{{Bovy}}{2015}]{bovy2015}
{Bovy} J.,  2015, \mn@doi [\apjs] {10.1088/0067-0049/216/2/29}, \href
  {https://ui.adsabs.harvard.edu/abs/2015ApJS..216...29B} {216, 29}

\bibitem[\protect\citeauthoryear{{Bovy} et~al.,}{{Bovy}
  et~al.}{2014}]{bovy2014}
{Bovy} J.,  et~al., 2014, \mn@doi [\apj] {10.1088/0004-637X/790/2/127}, \href
  {https://ui.adsabs.harvard.edu/abs/2014ApJ...790..127B} {790, 127}

\bibitem[\protect\citeauthoryear{{Bowen} \& {Vaughan}}{{Bowen} \&
  {Vaughan}}{1973}]{bowen1973}
{Bowen} I.~S.,  {Vaughan} A.~H. J.,  1973, \mn@doi [\ao]
  {10.1364/AO.12.001430}, \href
  {https://ui.adsabs.harvard.edu/abs/1973ApOpt..12.1430B} {12, 1430}

\bibitem[\protect\citeauthoryear{{Bragan{\c{c}}a} et~al.,}{{Bragan{\c{c}}a}
  et~al.}{2019}]{braganca2019}
{Bragan{\c{c}}a} G.~A.,  et~al., 2019, \mn@doi [\aap]
  {10.1051/0004-6361/201834554}, \href
  {https://ui.adsabs.harvard.edu/abs/2019A&A...625A.120B} {625, A120}

\bibitem[\protect\citeauthoryear{{Brunetti}, {Chiappini}  \&
  {Pfenniger}}{{Brunetti} et~al.}{2011}]{brunetti2011}
{Brunetti} M.,  {Chiappini} C.,   {Pfenniger} D.,  2011, \mn@doi [\aap]
  {10.1051/0004-6361/201117566}, \href
  {https://ui.adsabs.harvard.edu/abs/2011A&A...534A..75B} {534, A75}

\bibitem[\protect\citeauthoryear{{Buck}}{{Buck}}{2020}]{buck2020}
{Buck} T.,  2020, \mn@doi [\mnras] {10.1093/mnras/stz3289}, \href
  {https://ui.adsabs.harvard.edu/abs/2020MNRAS.491.5435B} {491, 5435}

\bibitem[\protect\citeauthoryear{{Buell}}{{Buell}}{2013}]{buell2013}
{Buell} J.~F.,  2013, \mn@doi [\mnras] {10.1093/mnras/sts211}, \href
  {https://ui.adsabs.harvard.edu/abs/2013MNRAS.428.2577B} {428, 2577}

\bibitem[\protect\citeauthoryear{{Calura} \& {Menci}}{{Calura} \&
  {Menci}}{2009}]{calura2009}
{Calura} F.,  {Menci} N.,  2009, \mn@doi [\mnras]
  {10.1111/j.1365-2966.2009.15440.x}, \href
  {https://ui.adsabs.harvard.edu/abs/2009MNRAS.400.1347C} {400, 1347}

\bibitem[\protect\citeauthoryear{{Chiappini}, {Matteucci}  \&
  {Gratton}}{{Chiappini} et~al.}{1997}]{chiappini1997}
{Chiappini} C.,  {Matteucci} F.,   {Gratton} R.,  1997, \mn@doi [\apj]
  {10.1086/303726}, \href
  {https://ui.adsabs.harvard.edu/abs/1997ApJ...477..765C} {477, 765}

\bibitem[\protect\citeauthoryear{{Daflon} \& {Cunha}}{{Daflon} \&
  {Cunha}}{2004}]{daflon2004}
{Daflon} S.,  {Cunha} K.,  2004, \mn@doi [\apj] {10.1086/425607}, \href
  {https://ui.adsabs.harvard.edu/abs/2004ApJ...617.1115D} {617, 1115}

\bibitem[\protect\citeauthoryear{{Daniel}, {Schaffner}, {McCluskey}, {Fiedler
  Kawaguchi}  \& {Loebman}}{{Daniel} et~al.}{2019}]{daniel2019}
{Daniel} K.~J.,  {Schaffner} D.~A.,  {McCluskey} F.,  {Fiedler Kawaguchi} C.,
  {Loebman} S.,  2019, \mn@doi [\apj] {10.3847/1538-4357/ab341a}, \href
  {https://ui.adsabs.harvard.edu/abs/2019ApJ...882..111D} {882, 111}

\bibitem[\protect\citeauthoryear{{Davies}, {Origlia}, {Kudritzki}, {Figer},
  {Rich}, {Najarro}, {Negueruela}  \& {Clark}}{{Davies}
  et~al.}{2009}]{davies2009}
{Davies} B.,  {Origlia} L.,  {Kudritzki} R.-P.,  {Figer} D.~F.,  {Rich} R.~M.,
  {Najarro} F.,  {Negueruela} I.,   {Clark} J.~S.,  2009, \mn@doi [\apj]
  {10.1088/0004-637X/696/2/2014}, \href
  {https://ui.adsabs.harvard.edu/abs/2009ApJ...696.2014D} {696, 2014}

\bibitem[\protect\citeauthoryear{{Di Matteo}, {Haywood}, {Combes}, {Semelin}
  \& {Snaith}}{{Di Matteo} et~al.}{2013}]{dimatteo2013}
{Di Matteo} P.,  {Haywood} M.,  {Combes} F.,  {Semelin} B.,   {Snaith} O.~N.,
  2013, \mn@doi [\aap] {10.1051/0004-6361/201220539}, \href
  {https://ui.adsabs.harvard.edu/abs/2013A&A...553A.102D} {553, A102}

\bibitem[\protect\citeauthoryear{{El-Badry}, {Wetzel}, {Geha}, {Hopkins},
  {Kere{\v{s}}}, {Chan}  \& {Faucher-Gigu{\`e}re}}{{El-Badry}
  et~al.}{2016}]{elbadry2016}
{El-Badry} K.,  {Wetzel} A.,  {Geha} M.,  {Hopkins} P.~F.,  {Kere{\v{s}}} D.,
  {Chan} T.~K.,   {Faucher-Gigu{\`e}re} C.-A.,  2016, \mn@doi [\apj]
  {10.3847/0004-637X/820/2/131}, \href
  {https://ui.adsabs.harvard.edu/abs/2016ApJ...820..131E} {820, 131}

\bibitem[\protect\citeauthoryear{{Feuillet} et~al.,}{{Feuillet}
  et~al.}{2018}]{feuillet2018}
{Feuillet} D.~K.,  et~al., 2018, \mn@doi [\mnras] {10.1093/mnras/sty779}, \href
  {https://ui.adsabs.harvard.edu/abs/2018MNRAS.477.2326F} {477, 2326}

\bibitem[\protect\citeauthoryear{{Feuillet}, {Frankel}, {Lind}, {Frinchaboy},
  {Garc{\'\i}a-Hern{\'a}ndez}, {Lane}, {Nitschelm}  \&
  {Roman-Lopes}}{{Feuillet} et~al.}{2019}]{feuillet2019}
{Feuillet} D.~K.,  {Frankel} N.,  {Lind} K.,  {Frinchaboy} P.~M.,
  {Garc{\'\i}a-Hern{\'a}ndez} D.~A.,  {Lane} R.~R.,  {Nitschelm} C.,
  {Roman-Lopes} A.,  2019, \mn@doi [\mnras] {10.1093/mnras/stz2221}, \href
  {https://ui.adsabs.harvard.edu/abs/2019MNRAS.489.1742F} {489, 1742}

\bibitem[\protect\citeauthoryear{{Frankel}, {Rix}, {Ting}, {Ness}  \&
  {Hogg}}{{Frankel} et~al.}{2018}]{frankel2018}
{Frankel} N.,  {Rix} H.-W.,  {Ting} Y.-S.,  {Ness} M.,   {Hogg} D.~W.,  2018,
  \mn@doi [\apj] {10.3847/1538-4357/aadba5}, \href
  {https://ui.adsabs.harvard.edu/abs/2018ApJ...865...96F} {865, 96}

\bibitem[\protect\citeauthoryear{{Frankel}, {Sanders}, {Ting}  \&
  {Rix}}{{Frankel} et~al.}{2020}]{frankel2020}
{Frankel} N.,  {Sanders} J.,  {Ting} Y.-S.,   {Rix} H.-W.,  2020, \mn@doi
  [\apj] {10.3847/1538-4357/ab910c}, \href
  {https://ui.adsabs.harvard.edu/abs/2020ApJ...896...15F} {896, 15}

\bibitem[\protect\citeauthoryear{{Garc{\'\i}a P{\'e}rez} et~al.,}{{Garc{\'\i}a
  P{\'e}rez} et~al.}{2016}]{garcia2016}
{Garc{\'\i}a P{\'e}rez} A.~E.,  et~al., 2016, \mn@doi [\aj]
  {10.3847/0004-6256/151/6/144}, \href
  {https://ui.adsabs.harvard.edu/abs/2016AJ....151..144G} {151, 144}

\bibitem[\protect\citeauthoryear{{Genovali} et~al.,}{{Genovali}
  et~al.}{2014}]{genovali2014}
{Genovali} K.,  et~al., 2014, \mn@doi [\aap] {10.1051/0004-6361/201323198},
  \href {https://ui.adsabs.harvard.edu/abs/2014A&A...566A..37G} {566, A37}

\bibitem[\protect\citeauthoryear{{Gesicki}, {Zijlstra}, {Hajduk}  \&
  {Szyszka}}{{Gesicki} et~al.}{2014}]{gesicki2014}
{Gesicki} K.,  {Zijlstra} A.~A.,  {Hajduk} M.,   {Szyszka} C.,  2014, \mn@doi
  [\aap] {10.1051/0004-6361/201118391}, \href
  {https://ui.adsabs.harvard.edu/abs/2014A&A...566A..48G} {566, A48}

\bibitem[\protect\citeauthoryear{{Grand}, {Kawata}  \& {Cropper}}{{Grand}
  et~al.}{2015}]{grand2015}
{Grand} R. J.~J.,  {Kawata} D.,   {Cropper} M.,  2015, \mn@doi [\mnras]
  {10.1093/mnras/stv016}, \href
  {https://ui.adsabs.harvard.edu/abs/2015MNRAS.447.4018G} {447, 4018}

\bibitem[\protect\citeauthoryear{{Grieco}, {Matteucci}, {Pipino}  \&
  {Cescutti}}{{Grieco} et~al.}{2012}]{grieco2012}
{Grieco} V.,  {Matteucci} F.,  {Pipino} A.,   {Cescutti} G.,  2012, \mn@doi
  [\aap] {10.1051/0004-6361/201219761}, \href
  {https://ui.adsabs.harvard.edu/abs/2012A&A...548A..60G} {548, A60}

\bibitem[\protect\citeauthoryear{{Griffith} et~al.,}{{Griffith}
  et~al.}{2021}]{griffith2020}
{Griffith} E.,  et~al., 2021, \mn@doi [\apj] {10.3847/1538-4357/abd6be}, \href
  {https://ui.adsabs.harvard.edu/abs/2021ApJ...909...77G} {909, 77}

\bibitem[\protect\citeauthoryear{{Gunn} et~al.,}{{Gunn}
  et~al.}{2006}]{gunn2006}
{Gunn} J.~E.,  et~al., 2006, \mn@doi [\aj] {10.1086/500975}, \href
  {https://ui.adsabs.harvard.edu/abs/2006AJ....131.2332G} {131, 2332}

\bibitem[\protect\citeauthoryear{{Halle}, {Di Matteo}, {Haywood}  \&
  {Combes}}{{Halle} et~al.}{2015}]{halle2015}
{Halle} A.,  {Di Matteo} P.,  {Haywood} M.,   {Combes} F.,  2015, \mn@doi
  [\aap] {10.1051/0004-6361/201525612}, \href
  {https://ui.adsabs.harvard.edu/abs/2015A&A...578A..58H} {578, A58}

\bibitem[\protect\citeauthoryear{{Hasselquist} et~al.,}{{Hasselquist}
  et~al.}{2019}]{hasselquist2019}
{Hasselquist} S.,  et~al., 2019, \mn@doi [\apj] {10.3847/1538-4357/aaf859},
  \href {https://ui.adsabs.harvard.edu/abs/2019ApJ...871..181H} {871, 181}

\bibitem[\protect\citeauthoryear{{Hasselquist} et~al.,}{{Hasselquist}
  et~al.}{2020}]{hasselquist2020}
{Hasselquist} S.,  et~al., 2020, \mn@doi [\apj] {10.3847/1538-4357/abaeee},
  \href {https://ui.adsabs.harvard.edu/abs/2020ApJ...901..109H} {901, 109}

\bibitem[\protect\citeauthoryear{{Hayden} et~al.,}{{Hayden}
  et~al.}{2015}]{hayden2015}
{Hayden} M.~R.,  et~al., 2015, \mn@doi [\apj] {10.1088/0004-637X/808/2/132},
  \href {https://ui.adsabs.harvard.edu/abs/2015ApJ...808..132H} {808, 132}

\bibitem[\protect\citeauthoryear{{Haywood}}{{Haywood}}{2008}]{haywood2008}
{Haywood} M.,  2008, \mn@doi [\mnras] {10.1111/j.1365-2966.2008.13395.x}, \href
  {https://ui.adsabs.harvard.edu/abs/2008MNRAS.388.1175H} {388, 1175}

\bibitem[\protect\citeauthoryear{{Haywood}, {Snaith}, {Lehnert}, {Di Matteo}
  \& {Khoperskov}}{{Haywood} et~al.}{2019}]{haywood2019}
{Haywood} M.,  {Snaith} O.,  {Lehnert} M.~D.,  {Di Matteo} P.,   {Khoperskov}
  S.,  2019, \mn@doi [\aap] {10.1051/0004-6361/201834155}, \href
  {https://ui.adsabs.harvard.edu/abs/2019A&A...625A.105H} {625, A105}

\bibitem[\protect\citeauthoryear{{Hill} et~al.,}{{Hill}
  et~al.}{2011}]{hill2011}
{Hill} V.,  et~al., 2011, \mn@doi [\aap] {10.1051/0004-6361/200913757}, \href
  {https://ui.adsabs.harvard.edu/abs/2011A&A...534A..80H} {534, A80}

\bibitem[\protect\citeauthoryear{{Ho} et~al.,}{{Ho} et~al.}{2017}]{ho2017}
{Ho} I.~T.,  et~al., 2017, \mn@doi [\apj] {10.3847/1538-4357/aa8460}, \href
  {https://ui.adsabs.harvard.edu/abs/2017ApJ...846...39H} {846, 39}

\bibitem[\protect\citeauthoryear{{Holtzman}, {Harrison}  \&
  {Coughlin}}{{Holtzman} et~al.}{2010}]{holtzman2010}
{Holtzman} J.~A.,  {Harrison} T.~E.,   {Coughlin} J.~L.,  2010, \mn@doi
  [Advances in Astronomy] {10.1155/2010/193086}, \href
  {https://ui.adsabs.harvard.edu/abs/2010AdAst2010E..46H} {2010, 193086}

\bibitem[\protect\citeauthoryear{{Jofre}}{{Jofre}}{2021}]{jofre2021}
{Jofre} P.,  2021, arXiv e-prints, \href
  {https://ui.adsabs.harvard.edu/abs/2021arXiv210616119J} {p. arXiv:2106.16119}

\bibitem[\protect\citeauthoryear{{Johnson} et~al.,}{{Johnson}
  et~al.}{2021}]{johnson2021}
{Johnson} J.~W.,  et~al., 2021, arXiv e-prints, \href
  {https://ui.adsabs.harvard.edu/abs/2021arXiv210309838J} {p. arXiv:2103.09838}

\bibitem[\protect\citeauthoryear{{J{\"o}nsson} et~al.,}{{J{\"o}nsson}
  et~al.}{2020}]{jonsson2020}
{J{\"o}nsson} H.,  et~al., 2020, \mn@doi [\aj] {10.3847/1538-3881/aba592},
  \href {https://ui.adsabs.harvard.edu/abs/2020AJ....160..120J} {160, 120}

\bibitem[\protect\citeauthoryear{{Khoperskov}, {Di Matteo}, {Haywood},
  {G{\'o}mez}  \& {Snaith}}{{Khoperskov} et~al.}{2020}]{khoperskov2020}
{Khoperskov} S.,  {Di Matteo} P.,  {Haywood} M.,  {G{\'o}mez} A.,   {Snaith}
  O.~N.,  2020, \mn@doi [\aap] {10.1051/0004-6361/201937188}, \href
  {https://ui.adsabs.harvard.edu/abs/2020A&A...638A.144K} {638, A144}

\bibitem[\protect\citeauthoryear{{Khoperskov}, {Haywood}, {Snaith}, {Di
  Matteo}, {Lehnert}, {Vasiliev}, {Naroenkov}  \& {Berczik}}{{Khoperskov}
  et~al.}{2021}]{khoperskov2021}
{Khoperskov} S.,  {Haywood} M.,  {Snaith} O.,  {Di Matteo} P.,  {Lehnert} M.,
  {Vasiliev} E.,  {Naroenkov} S.,   {Berczik} P.,  2021, \mn@doi [\mnras]
  {10.1093/mnras/staa3996}, \href
  {https://ui.adsabs.harvard.edu/abs/2021MNRAS.501.5176K} {501, 5176}

\bibitem[\protect\citeauthoryear{{Kreckel} et~al.,}{{Kreckel}
  et~al.}{2020}]{kreckel2020}
{Kreckel} K.,  et~al., 2020, \mn@doi [\mnras] {10.1093/mnras/staa2743}, \href
  {https://ui.adsabs.harvard.edu/abs/2020MNRAS.499..193K} {499, 193}

\bibitem[\protect\citeauthoryear{{Kubryk}, {Prantzos}  \&
  {Athanassoula}}{{Kubryk} et~al.}{2013}]{kubryk2013}
{Kubryk} M.,  {Prantzos} N.,   {Athanassoula} E.,  2013, \mn@doi [\mnras]
  {10.1093/mnras/stt1667}, \href
  {https://ui.adsabs.harvard.edu/abs/2013MNRAS.436.1479K} {436, 1479}

\bibitem[\protect\citeauthoryear{{Kubryk}, {Prantzos}  \&
  {Athanassoula}}{{Kubryk} et~al.}{2015}]{kubryk2015}
{Kubryk} M.,  {Prantzos} N.,   {Athanassoula} E.,  2015, \mn@doi [\aap]
  {10.1051/0004-6361/201424171}, \href
  {https://ui.adsabs.harvard.edu/abs/2015A&A...580A.126K} {580, A126}

\bibitem[\protect\citeauthoryear{{Leung} \& {Bovy}}{{Leung} \&
  {Bovy}}{2019}]{leung2019}
{Leung} H.~W.,  {Bovy} J.,  2019, \mn@doi [\mnras] {10.1093/mnras/sty3217},
  \href {https://ui.adsabs.harvard.edu/abs/2019MNRAS.483.3255L} {483, 3255}

\bibitem[\protect\citeauthoryear{{Li}, {Bresolin}  \& {Kennicutt}}{{Li}
  et~al.}{2013}]{li2013}
{Li} Y.,  {Bresolin} F.,   {Kennicutt} Robert~C. J.,  2013, \mn@doi [\apj]
  {10.1088/0004-637X/766/1/17}, \href
  {https://ui.adsabs.harvard.edu/abs/2013ApJ...766...17L} {766, 17}

\bibitem[\protect\citeauthoryear{{Lian} et~al.,}{{Lian}
  et~al.}{2020a}]{lian2020a}
{Lian} J.,  et~al., 2020a, \mn@doi [\mnras] {10.1093/mnras/staa867}, \href
  {https://ui.adsabs.harvard.edu/abs/2020MNRAS.494.2561L} {494, 2561}

\bibitem[\protect\citeauthoryear{{Lian} et~al.,}{{Lian}
  et~al.}{2020b}]{lian2020b}
{Lian} J.,  et~al., 2020b, \mn@doi [\mnras] {10.1093/mnras/staa2078}, \href
  {https://ui.adsabs.harvard.edu/abs/2020MNRAS.497.2371L} {497, 2371}

\bibitem[\protect\citeauthoryear{{Lian} et~al.,}{{Lian}
  et~al.}{2020c}]{lian2020c}
{Lian} J.,  et~al., 2020c, \mn@doi [\mnras] {10.1093/mnras/staa2205}, \href
  {https://ui.adsabs.harvard.edu/abs/2020MNRAS.497.3557L} {497, 3557}

\bibitem[\protect\citeauthoryear{{Lian} et~al.,}{{Lian}
  et~al.}{2021}]{lian2021}
{Lian} J.,  et~al., 2021, \mn@doi [\mnras] {10.1093/mnras/staa3256}, \href
  {https://ui.adsabs.harvard.edu/abs/2021MNRAS.500..282L} {500, 282}

\bibitem[\protect\citeauthoryear{{Lin}, {Dotter}, {Ting}  \& {Asplund}}{{Lin}
  et~al.}{2018}]{lin2018}
{Lin} J.,  {Dotter} A.,  {Ting} Y.-S.,   {Asplund} M.,  2018, \mn@doi [\mnras]
  {10.1093/mnras/sty709}, \href
  {https://ui.adsabs.harvard.edu/abs/2018MNRAS.477.2966L} {477, 2966}

\bibitem[\protect\citeauthoryear{{Lin} et~al.,}{{Lin} et~al.}{2019}]{lin2019}
{Lin} L.,  et~al., 2019, \mn@doi [\apj] {10.3847/1538-4357/aafa84}, \href
  {https://ui.adsabs.harvard.edu/abs/2019ApJ...872...50L} {872, 50}

\bibitem[\protect\citeauthoryear{{Loebman}, {Ro{\v{s}}kar}, {Debattista},
  {Ivezi{\'c}}, {Quinn}  \& {Wadsley}}{{Loebman} et~al.}{2011}]{loebman2011}
{Loebman} S.~R.,  {Ro{\v{s}}kar} R.,  {Debattista} V.~P.,  {Ivezi{\'c}}
  {\v{Z}}.,  {Quinn} T.~R.,   {Wadsley} J.,  2011, \mn@doi [\apj]
  {10.1088/0004-637X/737/1/8}, \href
  {https://ui.adsabs.harvard.edu/abs/2011ApJ...737....8L} {737, 8}

\bibitem[\protect\citeauthoryear{{Loebman}, {Debattista}, {Nidever}, {Hayden},
  {Holtzman}, {Clarke}, {Ro{\v{s}}kar}  \& {Valluri}}{{Loebman}
  et~al.}{2016}]{loebman2016}
{Loebman} S.~R.,  {Debattista} V.~P.,  {Nidever} D.~L.,  {Hayden} M.~R.,
  {Holtzman} J.~A.,  {Clarke} A.~J.,  {Ro{\v{s}}kar} R.,   {Valluri} M.,  2016,
  \mn@doi [\apjl] {10.3847/2041-8205/818/1/L6}, \href
  {https://ui.adsabs.harvard.edu/abs/2016ApJ...818L...6L} {818, L6}

\bibitem[\protect\citeauthoryear{{Luck} \& {Lambert}}{{Luck} \&
  {Lambert}}{2011}]{luck2011}
{Luck} R.~E.,  {Lambert} D.~L.,  2011, \mn@doi [\aj]
  {10.1088/0004-6256/142/4/136}, \href
  {https://ui.adsabs.harvard.edu/abs/2011AJ....142..136L} {142, 136}

\bibitem[\protect\citeauthoryear{{Lynden-Bell} \& {Kalnajs}}{{Lynden-Bell} \&
  {Kalnajs}}{1972}]{lynden-bell1972}
{Lynden-Bell} D.,  {Kalnajs} A.~J.,  1972, \mn@doi [\mnras]
  {10.1093/mnras/157.1.1}, \href
  {https://ui.adsabs.harvard.edu/abs/1972MNRAS.157....1L} {157, 1}

\bibitem[\protect\citeauthoryear{{Mackereth} et~al.,}{{Mackereth}
  et~al.}{2017}]{mackereth2017}
{Mackereth} J.~T.,  et~al., 2017, \mn@doi [\mnras] {10.1093/mnras/stx1774},
  \href {https://ui.adsabs.harvard.edu/abs/2017MNRAS.471.3057M} {471, 3057}

\bibitem[\protect\citeauthoryear{{Mackereth} et~al.,}{{Mackereth}
  et~al.}{2019}]{mackereth2019}
{Mackereth} J.~T.,  et~al., 2019, \mn@doi [\mnras] {10.1093/mnras/stz1521},
  \href {https://ui.adsabs.harvard.edu/abs/2019MNRAS.489..176M} {489, 176}

\bibitem[\protect\citeauthoryear{{Majewski} et~al.,}{{Majewski}
  et~al.}{2017}]{majewski2017}
{Majewski} S.~R.,  et~al., 2017, \mn@doi [\aj] {10.3847/1538-3881/aa784d},
  \href {https://ui.adsabs.harvard.edu/abs/2017AJ....154...94M} {154, 94}

\bibitem[\protect\citeauthoryear{{Minchev} \& {Famaey}}{{Minchev} \&
  {Famaey}}{2010}]{minchev2010}
{Minchev} I.,  {Famaey} B.,  2010, \mn@doi [\apj]
  {10.1088/0004-637X/722/1/112}, \href
  {https://ui.adsabs.harvard.edu/abs/2010ApJ...722..112M} {722, 112}

\bibitem[\protect\citeauthoryear{{Minchev}, {Famaey}, {Combes}, {Di Matteo},
  {Mouhcine}  \& {Wozniak}}{{Minchev} et~al.}{2011}]{minchev2011}
{Minchev} I.,  {Famaey} B.,  {Combes} F.,  {Di Matteo} P.,  {Mouhcine} M.,
  {Wozniak} H.,  2011, \mn@doi [\aap] {10.1051/0004-6361/201015139}, \href
  {https://ui.adsabs.harvard.edu/abs/2011A&A...527A.147M} {527, A147}

\bibitem[\protect\citeauthoryear{{Minchev}, {Chiappini}  \& {Martig}}{{Minchev}
  et~al.}{2013}]{minchev2013}
{Minchev} I.,  {Chiappini} C.,   {Martig} M.,  2013, \mn@doi [\aap]
  {10.1051/0004-6361/201220189}, \href
  {https://ui.adsabs.harvard.edu/abs/2013A&A...558A...9M} {558, A9}

\bibitem[\protect\citeauthoryear{{Minchev}, {Chiappini}  \& {Martig}}{{Minchev}
  et~al.}{2014}]{minchev2014}
{Minchev} I.,  {Chiappini} C.,   {Martig} M.,  2014, \mn@doi [\aap]
  {10.1051/0004-6361/201423487}, \href
  {https://ui.adsabs.harvard.edu/abs/2014A&A...572A..92M} {572, A92}

\bibitem[\protect\citeauthoryear{{Minchev} et~al.,}{{Minchev}
  et~al.}{2018}]{minchev2018}
{Minchev} I.,  et~al., 2018, \mn@doi [\mnras] {10.1093/mnras/sty2033}, \href
  {https://ui.adsabs.harvard.edu/abs/2018MNRAS.481.1645M} {481, 1645}

\bibitem[\protect\citeauthoryear{{Minniti} et~al.,}{{Minniti}
  et~al.}{2020}]{minniti2020}
{Minniti} J.~H.,  et~al., 2020, \mn@doi [\aap] {10.1051/0004-6361/202037575},
  \href {https://ui.adsabs.harvard.edu/abs/2020A&A...640A..92M} {640, A92}

\bibitem[\protect\citeauthoryear{{Nataf}}{{Nataf}}{2016}]{nataf2016}
{Nataf} D.~M.,  2016, \mn@doi [PASA] {10.1017/pasa.2015.38}, \href
  {https://ui.adsabs.harvard.edu/abs/2016PASA...33...23N} {33, e023}

\bibitem[\protect\citeauthoryear{{Nidever} et~al.,}{{Nidever}
  et~al.}{2015}]{nidever2015}
{Nidever} D.~L.,  et~al., 2015, \mn@doi [\aj] {10.1088/0004-6256/150/6/173},
  \href {https://ui.adsabs.harvard.edu/abs/2015AJ....150..173N} {150, 173}

\bibitem[\protect\citeauthoryear{{Nissen}, {Christensen-Dalsgaard},
  {Mosumgaard}, {Silva Aguirre}, {Spitoni}  \& {Verma}}{{Nissen}
  et~al.}{2020}]{nissen2020}
{Nissen} P.~E.,  {Christensen-Dalsgaard} J.,  {Mosumgaard} J.~R.,  {Silva
  Aguirre} V.,  {Spitoni} E.,   {Verma} K.,  2020, \mn@doi [\aap]
  {10.1051/0004-6361/202038300}, \href
  {https://ui.adsabs.harvard.edu/abs/2020A&A...640A..81N} {640, A81}

\bibitem[\protect\citeauthoryear{{Pinsonneault} et~al.,}{{Pinsonneault}
  et~al.}{2018}]{pinsonneault2018}
{Pinsonneault} M.~H.,  et~al., 2018, \mn@doi [\apjs]
  {10.3847/1538-4365/aaebfd}, \href
  {https://ui.adsabs.harvard.edu/abs/2018ApJS..239...32P} {239, 32}

\bibitem[\protect\citeauthoryear{{Queiroz} et~al.,}{{Queiroz}
  et~al.}{2020}]{queiroz2020}
{Queiroz} A.~B.~A.,  et~al., 2020, \mn@doi [\aap]
  {10.1051/0004-6361/201937364}, \href
  {https://ui.adsabs.harvard.edu/abs/2020A&A...638A..76Q} {638, A76}

\bibitem[\protect\citeauthoryear{{Quillen}, {Minchev}, {Bland-Hawthorn}  \&
  {Haywood}}{{Quillen} et~al.}{2009}]{quillen2009}
{Quillen} A.~C.,  {Minchev} I.,  {Bland-Hawthorn} J.,   {Haywood} M.,  2009,
  \mn@doi [\mnras] {10.1111/j.1365-2966.2009.15054.x}, \href
  {https://ui.adsabs.harvard.edu/abs/2009MNRAS.397.1599Q} {397, 1599}

\bibitem[\protect\citeauthoryear{{Renaud}, {Agertz}, {Andersson}, {Read},
  {Ryde}, {Bensby}, {Rey}  \& {Feuillet}}{{Renaud} et~al.}{2021}]{renaud2021}
{Renaud} F.,  {Agertz} O.,  {Andersson} E.~P.,  {Read} J.~I.,  {Ryde} N.,
  {Bensby} T.,  {Rey} M.~P.,   {Feuillet} D.~K.,  2021, \mn@doi [\mnras]
  {10.1093/mnras/stab543}, \href
  {https://ui.adsabs.harvard.edu/abs/2021MNRAS.503.5868R} {503, 5868}

\bibitem[\protect\citeauthoryear{{Rojas-Arriagada} et~al.,}{{Rojas-Arriagada}
  et~al.}{2017}]{rojas2017}
{Rojas-Arriagada} A.,  et~al., 2017, \mn@doi [\aap]
  {10.1051/0004-6361/201629160}, \href
  {https://ui.adsabs.harvard.edu/abs/2017A&A...601A.140R} {601, A140}

\bibitem[\protect\citeauthoryear{{Rojas-Arriagada}, {Zoccali}, {Schultheis},
  {Recio-Blanco}, {Zasowski}, {Minniti}, {J{\"o}nsson}  \&
  {Cohen}}{{Rojas-Arriagada} et~al.}{2019}]{rojas2019}
{Rojas-Arriagada} A.,  {Zoccali} M.,  {Schultheis} M.,  {Recio-Blanco} A.,
  {Zasowski} G.,  {Minniti} D.,  {J{\"o}nsson} H.,   {Cohen} R.~E.,  2019,
  \mn@doi [\aap] {10.1051/0004-6361/201834126}, \href
  {https://ui.adsabs.harvard.edu/abs/2019A&A...626A..16R} {626, A16}

\bibitem[\protect\citeauthoryear{{Rojas-Arriagada} et~al.,}{{Rojas-Arriagada}
  et~al.}{2020}]{rojas2020}
{Rojas-Arriagada} A.,  et~al., 2020, \mn@doi [\mnras] {10.1093/mnras/staa2807},
  \href {https://ui.adsabs.harvard.edu/abs/2020MNRAS.tmp.2647R} {}

\bibitem[\protect\citeauthoryear{{Ro{\v{s}}kar}, {Debattista}, {Stinson},
  {Quinn}, {Kaufmann}  \& {Wadsley}}{{Ro{\v{s}}kar}
  et~al.}{2008a}]{roskar2008b}
{Ro{\v{s}}kar} R.,  {Debattista} V.~P.,  {Stinson} G.~S.,  {Quinn} T.~R.,
  {Kaufmann} T.,   {Wadsley} J.,  2008a, \mn@doi [\apjl] {10.1086/586734},
  \href {https://ui.adsabs.harvard.edu/abs/2008ApJ...675L..65R} {675, L65}

\bibitem[\protect\citeauthoryear{{Ro{\v{s}}kar}, {Debattista}, {Quinn},
  {Stinson}  \& {Wadsley}}{{Ro{\v{s}}kar} et~al.}{2008b}]{roskar2008a}
{Ro{\v{s}}kar} R.,  {Debattista} V.~P.,  {Quinn} T.~R.,  {Stinson} G.~S.,
  {Wadsley} J.,  2008b, \mn@doi [\apjl] {10.1086/592231}, \href
  {https://ui.adsabs.harvard.edu/abs/2008ApJ...684L..79R} {684, L79}

\bibitem[\protect\citeauthoryear{{Ruiz-Lara} et~al.,}{{Ruiz-Lara}
  et~al.}{2017}]{ruiz-lara2017}
{Ruiz-Lara} T.,  et~al., 2017, \mn@doi [\aap] {10.1051/0004-6361/201730705},
  \href {https://ui.adsabs.harvard.edu/abs/2017A&A...604A...4R} {604, A4}

\bibitem[\protect\citeauthoryear{{Sahlholdt}, {Feltzing}  \&
  {Feuillet}}{{Sahlholdt} et~al.}{2021}]{sahlholdt2021}
{Sahlholdt} C.~L.,  {Feltzing} S.,   {Feuillet} D.~K.,  2021, arXiv e-prints,
  \href {https://ui.adsabs.harvard.edu/abs/2021arXiv211208218S} {p.
  arXiv:2112.08218}

\bibitem[\protect\citeauthoryear{{S{\'a}nchez} et~al.,}{{S{\'a}nchez}
  et~al.}{2015}]{sanchez2015}
{S{\'a}nchez} S.~F.,  et~al., 2015, \mn@doi [\aap]
  {10.1051/0004-6361/201424950}, \href
  {https://ui.adsabs.harvard.edu/abs/2015A&A...573A.105S} {573, A105}

\bibitem[\protect\citeauthoryear{{Schaefer} et~al.,}{{Schaefer}
  et~al.}{2017}]{schaefer2017}
{Schaefer} A.~L.,  et~al., 2017, \mn@doi [\mnras] {10.1093/mnras/stw2289},
  \href {https://ui.adsabs.harvard.edu/abs/2017MNRAS.464..121S} {464, 121}

\bibitem[\protect\citeauthoryear{{Sch{\"o}nrich} \& {Binney}}{{Sch{\"o}nrich}
  \& {Binney}}{2009}]{schonrich2009}
{Sch{\"o}nrich} R.,  {Binney} J.,  2009, \mn@doi [\mnras]
  {10.1111/j.1365-2966.2009.14750.x}, \href
  {https://ui.adsabs.harvard.edu/abs/2009MNRAS.396..203S} {396, 203}

\bibitem[\protect\citeauthoryear{{Schultheis} et~al.,}{{Schultheis}
  et~al.}{2017}]{schultheis2017}
{Schultheis} M.,  et~al., 2017, \mn@doi [\aap] {10.1051/0004-6361/201630154},
  \href {https://ui.adsabs.harvard.edu/abs/2017A&A...600A..14S} {600, A14}

\bibitem[\protect\citeauthoryear{{Sellwood} \& {Binney}}{{Sellwood} \&
  {Binney}}{2002}]{sellwood2002}
{Sellwood} J.~A.,  {Binney} J.~J.,  2002, \mn@doi [\mnras]
  {10.1046/j.1365-8711.2002.05806.x}, \href
  {https://ui.adsabs.harvard.edu/abs/2002MNRAS.336..785S} {336, 785}

\bibitem[\protect\citeauthoryear{{Sharma}, {Hayden}  \&
  {Bland-Hawthorn}}{{Sharma} et~al.}{2020}]{sharma2020}
{Sharma} S.,  {Hayden} M.~R.,   {Bland-Hawthorn} J.,  2020, arXiv e-prints,
  \href {https://ui.adsabs.harvard.edu/abs/2020arXiv200503646S} {p.
  arXiv:2005.03646}

\bibitem[\protect\citeauthoryear{{Shetrone} et~al.,}{{Shetrone}
  et~al.}{2019}]{shetrone2019}
{Shetrone} M.,  et~al., 2019, \mn@doi [\apj] {10.3847/1538-4357/aaff66}, \href
  {https://ui.adsabs.harvard.edu/abs/2019ApJ...872..137S} {872, 137}

\bibitem[\protect\citeauthoryear{{Silva Aguirre} et~al.,}{{Silva Aguirre}
  et~al.}{2018}]{silva2018}
{Silva Aguirre} V.,  et~al., 2018, \mn@doi [\mnras] {10.1093/mnras/sty150},
  \href {https://ui.adsabs.harvard.edu/abs/2018MNRAS.475.5487S} {475, 5487}

\bibitem[\protect\citeauthoryear{{Smith} et~al.,}{{Smith}
  et~al.}{2021}]{smith2021}
{Smith} V.~V.,  et~al., 2021, \mn@doi [\aj] {10.3847/1538-3881/abefdc}, \href
  {https://ui.adsabs.harvard.edu/abs/2021AJ....161..254S} {161, 254}

\bibitem[\protect\citeauthoryear{{Spitoni}, {Silva Aguirre}, {Matteucci},
  {Calura}  \& {Grisoni}}{{Spitoni} et~al.}{2019}]{spitoni2019}
{Spitoni} E.,  {Silva Aguirre} V.,  {Matteucci} F.,  {Calura} F.,   {Grisoni}
  V.,  2019, \mn@doi [\aap] {10.1051/0004-6361/201834188}, \href
  {https://ui.adsabs.harvard.edu/abs/2019A&A...623A..60S} {623, A60}

\bibitem[\protect\citeauthoryear{{Spitoni}, {Verma}, {Silva Aguirre}  \&
  {Calura}}{{Spitoni} et~al.}{2020}]{spitoni2020}
{Spitoni} E.,  {Verma} K.,  {Silva Aguirre} V.,   {Calura} F.,  2020, \mn@doi
  [\aap] {10.1051/0004-6361/201937275}, \href
  {https://ui.adsabs.harvard.edu/abs/2020A&A...635A..58S} {635, A58}

\bibitem[\protect\citeauthoryear{{Stanghellini} \& {Haywood}}{{Stanghellini} \&
  {Haywood}}{2018}]{stanghellini2018}
{Stanghellini} L.,  {Haywood} M.,  2018, \mn@doi [\apj]
  {10.3847/1538-4357/aacaf8}, \href
  {https://ui.adsabs.harvard.edu/abs/2018ApJ...862...45S} {862, 45}

\bibitem[\protect\citeauthoryear{{Vincenzo} \& {Kobayashi}}{{Vincenzo} \&
  {Kobayashi}}{2020}]{vincenzo2020}
{Vincenzo} F.,  {Kobayashi} C.,  2020, \mn@doi [\mnras]
  {10.1093/mnras/staa1451}, \href
  {https://ui.adsabs.harvard.edu/abs/2020MNRAS.496...80V} {496, 80}

\bibitem[\protect\citeauthoryear{{Wenger}, {Balser}, {Anderson}  \&
  {Bania}}{{Wenger} et~al.}{2019}]{wenger2019}
{Wenger} T.~V.,  {Balser} D.~S.,  {Anderson} L.~D.,   {Bania} T.~M.,  2019,
  \mn@doi [\apj] {10.3847/1538-4357/ab53d3}, \href
  {https://ui.adsabs.harvard.edu/abs/2019ApJ...887..114W} {887, 114}

\bibitem[\protect\citeauthoryear{{Wilson} et~al.,}{{Wilson}
  et~al.}{2019}]{wilson2019}
{Wilson} J.~C.,  et~al., 2019, \mn@doi [\pasp] {10.1088/1538-3873/ab0075},
  \href {https://ui.adsabs.harvard.edu/abs/2019PASP..131e5001W} {131, 055001}

\bibitem[\protect\citeauthoryear{{Wu} et~al.,}{{Wu} et~al.}{2018}]{wu2018}
{Wu} Y.,  et~al., 2018, \mn@doi [\mnras] {10.1093/mnras/stx3296}, \href
  {https://ui.adsabs.harvard.edu/abs/2018MNRAS.475.3633W} {475, 3633}

\bibitem[\protect\citeauthoryear{{Wu} et~al.,}{{Wu} et~al.}{2019}]{wu2019}
{Wu} Y.,  et~al., 2019, \mn@doi [\mnras] {10.1093/mnras/stz256}, \href
  {https://ui.adsabs.harvard.edu/abs/2019MNRAS.484.5315W} {484, 5315}

\bibitem[\protect\citeauthoryear{{Xiang} et~al.,}{{Xiang}
  et~al.}{2017}]{xiang2017}
{Xiang} M.,  et~al., 2017, \mn@doi [\apjs] {10.3847/1538-4365/aa80e4}, \href
  {https://ui.adsabs.harvard.edu/abs/2017ApJS..232....2X} {232, 2}

\bibitem[\protect\citeauthoryear{{Yoachim}, {Ro{\v{s}}kar}  \&
  {Debattista}}{{Yoachim} et~al.}{2012}]{yoachim2012}
{Yoachim} P.,  {Ro{\v{s}}kar} R.,   {Debattista} V.~P.,  2012, \mn@doi [\apj]
  {10.1088/0004-637X/752/2/97}, \href
  {https://ui.adsabs.harvard.edu/abs/2012ApJ...752...97Y} {752, 97}

\bibitem[\protect\citeauthoryear{{Zasowski} et~al.,}{{Zasowski}
  et~al.}{2013}]{zasowski2013}
{Zasowski} G.,  et~al., 2013, \mn@doi [\aj] {10.1088/0004-6256/146/4/81}, \href
  {https://ui.adsabs.harvard.edu/abs/2013AJ....146...81Z} {146, 81}

\bibitem[\protect\citeauthoryear{{Zasowski} et~al.,}{{Zasowski}
  et~al.}{2017}]{zasowski2017}
{Zasowski} G.,  et~al., 2017, \mn@doi [\aj] {10.3847/1538-3881/aa8df9}, \href
  {https://ui.adsabs.harvard.edu/abs/2017AJ....154..198Z} {154, 198}

\bibitem[\protect\citeauthoryear{{Zhang}, {Chen}  \& {Zhao}}{{Zhang}
  et~al.}{2021}]{zhang2021}
{Zhang} H.,  {Chen} Y.,   {Zhao} G.,  2021, arXiv e-prints, \href
  {https://ui.adsabs.harvard.edu/abs/2021arXiv210612841Z} {p. arXiv:2106.12841}

\bibitem[\protect\citeauthoryear{{Zoccali} et~al.,}{{Zoccali}
  et~al.}{2003}]{zoccali2003}
{Zoccali} M.,  et~al., 2003, \mn@doi [\aap] {10.1051/0004-6361:20021604}, \href
  {https://ui.adsabs.harvard.edu/abs/2003A&A...399..931Z} {399, 931}

\makeatother
\end{thebibliography}

\appendix 
\section{Testing the effect of stellar parameters and choice of Galactocentric radius}

\begin{figure}
	\centering
	\includegraphics[width=16.5cm,viewport=10 10 1100 500,clip]{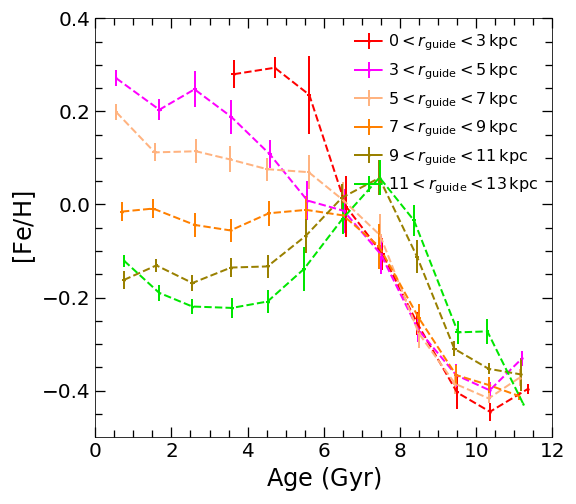}
	\caption{Same as Fig.~\ref{age-feh-median} but showing the sample after resampling in log($g$)-$T_{\rm eff}$ to the radial bin at $3<r_{\rm guide}<5$~kpc.   
	} 
	\label{logg-check}
\end{figure} 

\begin{figure}
	\centering
	\includegraphics[width=16.5cm,viewport=10 10 1100 500,clip]{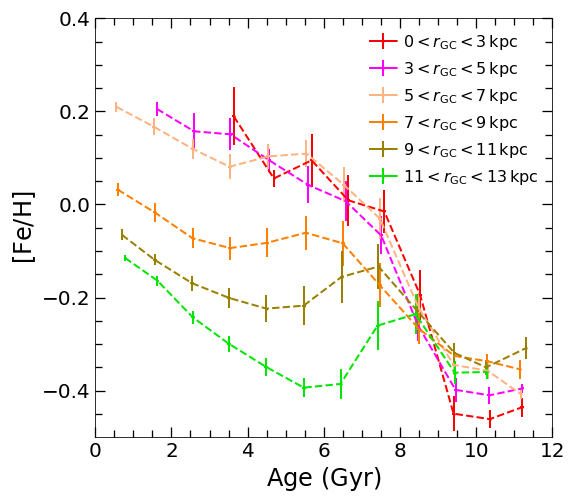}
	\caption{Same as Fig.~\ref{age-feh-median} but showing the sample selected using Galactocentric radius.    
	} 
	\label{radius-check}
\end{figure} 

\end{document}